\documentclass[%
 reprint,
showpacs,preprintnumbers,
 amsmath,amssymb,
 aps,
prc,
superscriptaddress]{revtex4-1}

\usepackage{graphicx}
\usepackage{dcolumn}
\usepackage{bm}

\begin{document}


\title{Modified  MIT bag Models - part I:\\ Thermodynamic consistency, stability windows and symmetry group}

\author{Luiz L. Lopes}
\email{llopes@cefetmg.br}
\affiliation{Centro Federal de Educa\c c\~ao  Tecnol\'ogica de
  Minas Gerais Campus VIII, CEP 37.022-560, Varginha, MG, Brasil}
\author{Carline Biesdorf}
\author{D\'ebora P. Menezes}
\affiliation{%
 Departamento de Fisica, CFM - Universidade Federal de Santa Catarina;  C.P. 476, CEP 88.040-900, Florian\'opolis, SC, Brasil 
}%

\date{\today}

\begin{abstract}
In this work we study different variations of the MIT bag model. We
start with the so called non-ideal bag model and discuss it in detail.
Then we implement a vector interaction in the MIT bag model that
simulates a meson exchange interaction and  fix the quark-meson coupling
 constants via symmetry group theory. At  the end we propose an
 original model, inspired by the Boguta-Bodmer models, which allows us
 to control the repulsion interaction at high densities. For each
 version of the model we obtain a stability window as predicted by the
 Bodmer-Witten conjecture and discuss its  thermodynamic consistency.

\end{abstract}


\maketitle

\section{Introduction}

According to the quantum chromodynamics (QCD) phase diagram,
deconfined quark matter existed in the early universe when the
temperature was very high. The region, where it is foreseen is called
quark-gluon plasma (QGP). At zero or low temperatures, however,
deconfined quark matter may be present in the core of massive compact
objects like neutron stars, as quark matter becomes energetically favorable over hadronic matter.

Moreover, Bodmer and Witten proposed that the ordinary matter we know, composed of protons and neutrons may be only meta-stable. The true ground state of strongly interacting matter would therefore  consist of strange matter (SM), which in turn is composed of deconfined up, down and strange quarks~\cite{Bod,Witten}.  
If this is true, as soon as the core of the star converts to the quark phase, the entire star will convert into what is called a strange star
in a finite amount of time~\cite{Olinto87}.

Also, it is worth pointing out that if the quark-quark interaction is attractive in any channel, then the true ground state of the system may not be the naked Fermi surface, but rather a complicated coherent state of particle and hole pairs - {\it Cooper pairs}, generally called {\it color superconductivity}. The quark pairs play the same role here as the Higgs particle does in
the standard model; the color-superconducting phase can be thought of as the
spontaneously broken (as opposed to confined) phase of QCD. A detailed study of the color-superconducting phase is beyond the scope of this work, but the reader can see more details in ref.~\cite{Alford1,Alford2}.

There have been some models used to approach the SM hypothesis, 
the first of them \cite{MITL} being the original MIT bag model \cite{MIT_original}. 
A modified version of this original model, called a non-ideal bag model \cite{MR1,MR2} has also been used to test this hypothesis. 
Furthermore, the quark-mass density dependent (QMDD) \cite{QMDD} 
and the Nambu-Jona-Lasinio (NJL) \cite{NJL1,NJL2,NJL3,NJL4} models have also been used to this same purpose. 
However, they present some issues:
the NJL model does not satisfy the Bodmer-Witten conjecture \cite{Buballa2005} while the QMDD models present some thermodynamic inconsistencies. Although the original model \cite{QMDD} was revisited several times (see \cite{James2013}, \cite{James_Veronica} and references therein), the problem persisted. More recently,
another QMDD model without this defect was introduced \cite{Peng2000},\cite{Xia2014} and recently also revisited in \cite{Betania}. 
Regarding the original version of the MIT bag model, it was shown it
is not able to reproduce massive stars as the ones recently detected \cite{James2013}.

In order to better describe quark stars, we would like to have a quark matter model that could reconcile thermodynamic consistency with the Bodmer-Witten conjecture, i.e, the SM being the true ground of the matter, with the existence of massive stars recently detected; this is the main motivation for this work. Towards this quest, we first revisit the original MIT bag model in section II and the non-ideal-bag model in section III and discuss their main caveats. To circumvent the problems we encounter, a improved version of the vector MIT bag model, introduced in ~\cite{Ro1,Ro2,Ro3,Klan1,Klan2,weiwei}, is proposed in section IV with the help of symmetry group techniques. Finally, inspired by the Boguta-Bodmer \cite{Boguta,Bodmer} non-linear terms in quantum hadrodynamics, a self-interacting vector field is introduced in section V. These non-linear terms mimic the Dirac sea contribution, which is absent in mean field approximation~~\cite{Serot97}. Our work ends in  section VI with a quick discussion about the tidal deformation for quark stars and the comparison with the GW170817 event constraints. Finally, the overall conclusions are presented in section VII.

\section{Original MIT bag model}

The MIT bag model considers that each baryon is composed of three
non-interacting quarks inside a bag. The bag, in its turn, corresponds to
an infinity potential that confines the quarks. In this simple model
the quarks are free inside the bag and are forbidden to reach its
exterior. All the information about the strong force relies on the
bag pressure value, which mimics the vacuum pressure. The MIT bag model Lagrangian
reads~\cite{MITL}:

\begin{equation}
\mathcal{L} = \sum_{u,d,s}\{ \bar{\psi}_q  [ i\gamma^{\mu} \partial_\mu - m_q ]\psi_q - B \}\Theta(\bar{\psi_q}\psi_q), \label{e1}
\end{equation}   
where $m_q$ is the mass of the quark $q$ of flavor u, d or s, $\psi_q$ is the Dirac quark field, B is the constant vacuum pressure and $\Theta(\bar{\psi_q}\psi_q)$ is
the Heaviside step function to assure that the quarks exist only confined to the bag. 

Applying the Euler-Lagrange equations for the quarks, we have:

\begin{equation}
[i\gamma^{\mu} \partial_\mu - m_q ]\psi_q =0 , \label{e2}
\end{equation}
which gives us a energy eigenvalue for the quark $q$:

\begin{equation}
E_q =  \sqrt{m_q^2 + k^2},  \label{e3}
\end{equation}
where $k$ is the momentum.

Quarks are fermions with spin 1/2. Hence, the number density and the energy density of the quark matter can be obtained via Fermi-Dirac distribution~\cite{Greiner}.
 At zero temperature approximation the Fermi-Dirac distribution becomes the Heaviside step function, and the energy eigenvalue matches the 
chemical potential, $E_q = \mu_q$. Therefore the quark number density is:

\begin{equation}
n_q = 2N_c \int_{0}^{k_f} \frac{d^3k}{(2\pi)^3} \Theta(\mu_q - E_q), = N_c\frac{k_{f}^3}{3\pi^2} \label{e4a}
\end{equation} 
while the energy density is:

\begin{equation}
\epsilon_q = \frac{N_c}{\pi^2}\int_{0}^{k_f} E_q k^2 dk , \label{e4}
\end{equation}
where $N_c$ is the number of colors and $k_f$ is the Fermi momentum. On the other hand the bag contribution to the energy density is easily obtained through the Hamiltonian:
$\mathcal{H} = -\langle \mathcal{L} \rangle$ = B.  Then, the total energy density is here the sum over the three lightest quarks plus the bag
contribution. The pressure $p$ can be obtained via thermodynamic relations:

\begin{equation}
\epsilon  = \sum_q \epsilon_q + B \quad \mbox{and} \quad  p = \sum_q n_q\mu_q - \epsilon = -\Omega  \label{e5}
\end{equation}
where $n_q$ is the number density, $\mu_q$ is the chemical potential and $\Omega$ is the thermodynamic potential.
It is also useful to write down the pressure (as well the
thermodynamic  potential) as the derivative of the energy
with respect to the volume:

\begin{equation}
p_q = -\Omega_q = - \bigg ( \frac{\partial E_q}{\partial V} \bigg )_{T} = \frac{N_c}{3\pi^2} \int_0^{k_f} \frac{k^4 dk}{E_q} \label{e6}
\end{equation}
and the total pressure is the sum of the pressure of each quark flavor plus the contribution of the bag from the Lagrangian:
p =  $\langle \mathcal{L} \rangle$. Therefore we obtain:

\begin{equation}
p = \sum_q p_q - B \label{e7} .
\end{equation}

Now, the bag constant is not totally arbitrary. In fact $B$ needs to
assure that a two-flavored quark matter is unstable and that it is not the ground state of the hadrons, i.e., its energy per baryon must be higher than $930$ MeV at zero pressure, otherwise the protons and neutrons
would decay into $u$ and $d$ quarks. On the other hand, if the
Bodmer-Witten conjecture~\cite{Bod,Witten}, that states that the true ground state of the hadronic matter is not consisted of baryons, but strange matter consisting of $\mu_u = \mu_d = \mu_s$, is true,
the three-flavored quark matter needs to be stable (energy per baryon
lower than $930$ MeV), while the two-flavored quark matter is unstable. 
Therefore, the bag pressure value $B$ can only assume a range of values,
known as the stability window~\cite{Bod, Witten, James2013}. 
These values depend on the quark masses. In this work, we assume that
the masses of the $u$ and $d$ quarks are 4 MeV. While in the past the mass of the $s$
quark was very ambiguous, today  it is  known to be around 95 MeV, more precisely, 93$^{+11}_{-5}$
MeV~\cite{smass}.  With these values, we display the range of $B$ for
the stability window in Table \ref{T1}. Slightly different values are
obtained if different quark masses are chosen, as seen in \cite{James2013}. Although the minimum value of $B$ remains the same, no matter the mass of the quark s, the upper limit varies from $B^{1/4}=162$~ MeV for $m_s=0$ to $B^{1/4}=148$~ MeV for $m_s=240$~MeV, i.e., no stable strange matter is possible if $m_s > 240$~MeV.

\begin{table}[h]
\begin{center}
\begin{tabular}{|c|c|c|}
\hline
 ~ MIT~&~Min. $B^{1/4}$~&~Max. $B^{1/4}$~ \\
\hline
 -    & 148 MeV & 159 MeV     \\
 \hline
 \end{tabular} 
\caption{Stability window for the MIT bag model.} 
\label{T1}
\end{center}
\end{table}

\section{non-ideal bag model}

The non-ideal bag model, presented in ref.~\cite{MR1,MR2}, is an empirical correction to the thermodynamic potential - $\Omega$ -
(or directly to the pressure as in ref.~\cite{Ellis92}) 
to match QCD correction of $O(\alpha_s)$ where $\alpha_s$ is the strong coupling constant~\cite{MR1}. In this model, the deviation from the original MIT bag model comes from an adimensional
parameter $a_4$, where $a_4 = (1 - 2\alpha_s/\pi)$. Although in the non-ideal bag model this term is included in an $ad~hoc$ way to the thermodynamic potential to reproduce some results coming 
from lattice QCD, it can be derived both from perturbative QCD~\cite{cold}, and from the hard dense loop (HDL) approximation~\cite{HDL}.
 When $a_4$ = 1, the original MIT bag model is recovered.
For massless quarks the thermodynamic potential reads~\cite{MR1,MR2,Ellis92}:

\begin{equation}
\Omega = -a_4\frac{N_f}{4\pi^2}\mu^4  + B_{eff} , \label{R1}
\end{equation} 
where, $\mu$  is the chemical potential of massless quarks, $N_f$ is the number of massless quarks,
 and $B_{eff}$ is an effective bag pressure value. With this model, the energy and
the pressure for each of the massless quarks are given by~\cite{MR2}:

\begin{eqnarray}
p = a_4\frac{1}{4\pi^2}\mu^4  - B_{eff}, \nonumber  \\
\epsilon = a_4 \frac{3}{4\pi^2}\mu^4  + B_{eff} , \label{R2} \\
\epsilon  = 3p + 4B_{eff}. \nonumber
\end{eqnarray}

Although this model does not modify the equation of state
$(p(\epsilon))$, it was proposed so that the parameter $a_4$ could change the relation
between the pressure and the chemical potential - $p(\mu)$.
 We discuss here some subtleties of this model.

As this model, in principle, is not constructed at the
Lagrangian level, it is hard to understand what kind of interaction
could produce such behavior.  Also, this model can present thermodynamic inconsistencies if we do not proceed carefully. To maintain the consistency of the model, the number density
needs to be scaled too, as in~\cite{Buballa2005}:

\begin{equation}
 n_q = a_4 \bigg ( N_c \frac{k_f^3}{3\pi^2} \bigg ) = a_4n_q \label{R5a} .
\end{equation}

Therefore, here  the model is thermodynamically consistent. 
In addition, assuming only Eq.~(\ref{R5a}), all Eqs.~(\ref{R2})
come as a consequence for massless quarks in the original MIT model. 
Therefore, the non-ideal bag model is actually  a modification in
the quark number, instead of a correction in the MIT model itself.
On other hand, the price we have to pay is the use a modification of  the Fermi-Dirac distribution~\cite{Greiner}. This modification comes from a renormalization at (4 - 2$\epsilon$)
dimensions in perturbative QCD~\cite{cold}.

To summarize, the non-ideal bag model can be considered as an empirical
correction from QCD. With this in mind we construct a stability
window for the non-ideal bag model, but using 
massive quarks, with the same masses as used previously, in the original MIT model.
Therefore, for each flavor, we have the following EoS:

\begin{eqnarray}
\epsilon_q = a_4\frac{N_c}{\pi^2}\int_0^{k_f}E_q k^2 dk , \label{R5}\\
p_q = a_4\frac{N_c}{3\pi^2}\int_0^{k_f} \frac{k^4 dk}{E_q} , \nonumber \\
n_q = a_4 N_c \frac{k_f^3}{3\pi^2}
\end{eqnarray}
and

\begin{equation}
\epsilon = \sum_q \epsilon_q + B_{eff} \quad \mbox{and} \quad p =  \sum_q p_q - B_{eff}. \label{R6}
\end{equation}

\begin{figure}[ht] 
\begin{centering}
 \includegraphics[angle=270,
width=0.4\textwidth]{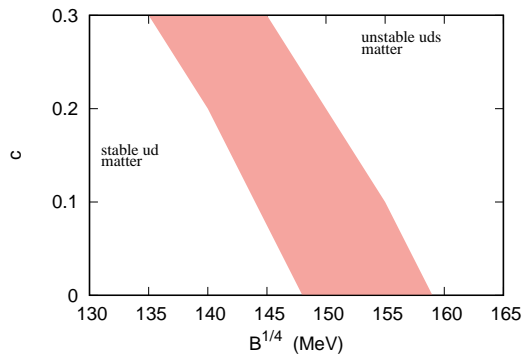}
\caption{(Color online) Stability window for the non-ideal bag model. If $c = 0$ we recover the original MIT bag model.} \label{F1}
\end{centering}
\end{figure}

Now let's redefine $a_4$ as $a_4 = 1- c$ as done in
ref.~\cite{MR2}. In this case, if $c = 0$ we recover the original MIT
bag model. As we increase $c$ we deviate from the traditional
model. We construct a stability window for values varying from 0 to
0.3, which coincides with the values used in ref.~\cite{MR1,MR2}. The
stability window is displayed in Fig.~\ref{F1} and the corresponding
values are presented 
in Tab.~\ref{T2}.

\begin{table}[h]
\begin{center}
\begin{tabular}{|c|c|c|}
\hline
 ~ c~&~Min. $B^{1/4}$~&~Max. $B^{1/4}$~ \\
\hline
 0.0    & 148 MeV & 159 MeV     \\
 0.1    & 144 MeV & 155 MeV     \\
 0.2    & 140 MeV & 150 MeV     \\
 0.3    & 135 MeV & 145 MeV     \\
 \hline
 \end{tabular} 
\caption{Stability windows for the non-ideal bag model.} 
\label{T2}
\end{center}
\end{table}

As we can see, as we increase the value of $c$, we displace 
the stability window within almost the same range. Increasing $c$ from
0 to 0.3 displaces the minimum value of $B^{1/4}$ from 148 MeV to 135 MeV.
On other hand, the range is almost the same, 11 MeV for $c = 0$ to
10 MeV for $c = 0.3$.

It is worth keeping in mind that the regions outside the stability window
on the left and on the right have very different meaning and
consequences. The $B^{1/4}$ 
cannot be lower than the minimum value presented
in Tab.~\ref{T2} because this would imply that two-flavored quark matter is stable
and our known universe composed of protons and neutron would no longer
exist.  But $B^{1/4}$ can be higher than the maximum value. This 
implies that the strange quark matter is not the ground state of the matter.
In this case deconfined quark matter can only be present in the core 
of massive hybrid stars, instead of forming quark stars.

The free parameter $c=2\alpha_s/\pi$ is here chosen below 0.3 to match the values used
in ref.~\cite{MR1,MR2}. However it is worth pointing out that $\alpha_s$ allows values of $c \approx 2$, as can be seen in ref.~\cite{Deur_et_al}, which would produce negative $a_4$ values. Therefore, the non-ideal bag model also presents some limitations. Nevertheless, our aim is to compare this model to the original MIT bag model and its modifications suggested in the next sections. For this purpose, we choose the $c$ values used in ref.~\cite{MR1,MR2}.

  \begin{figure}[ht] 
\begin{centering}
 \includegraphics[angle=270,
width=0.4\textwidth]{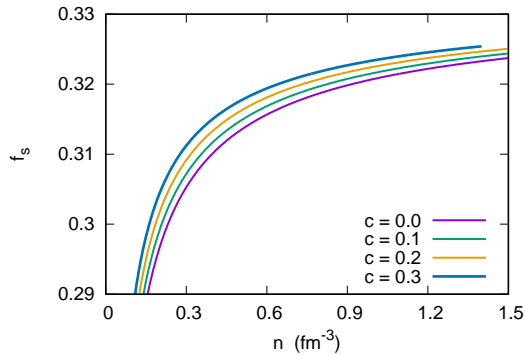}
\caption{(Color online) Strangeness fraction for different values of $c$. Increasing $c$ causes an increase of $f_s$.} \label{F1a}
\end{centering}
\end{figure}

\begin{figure*}[ht]
\begin{tabular}{cc}
\includegraphics[width=5.6cm,height=7.0cm,angle=270]{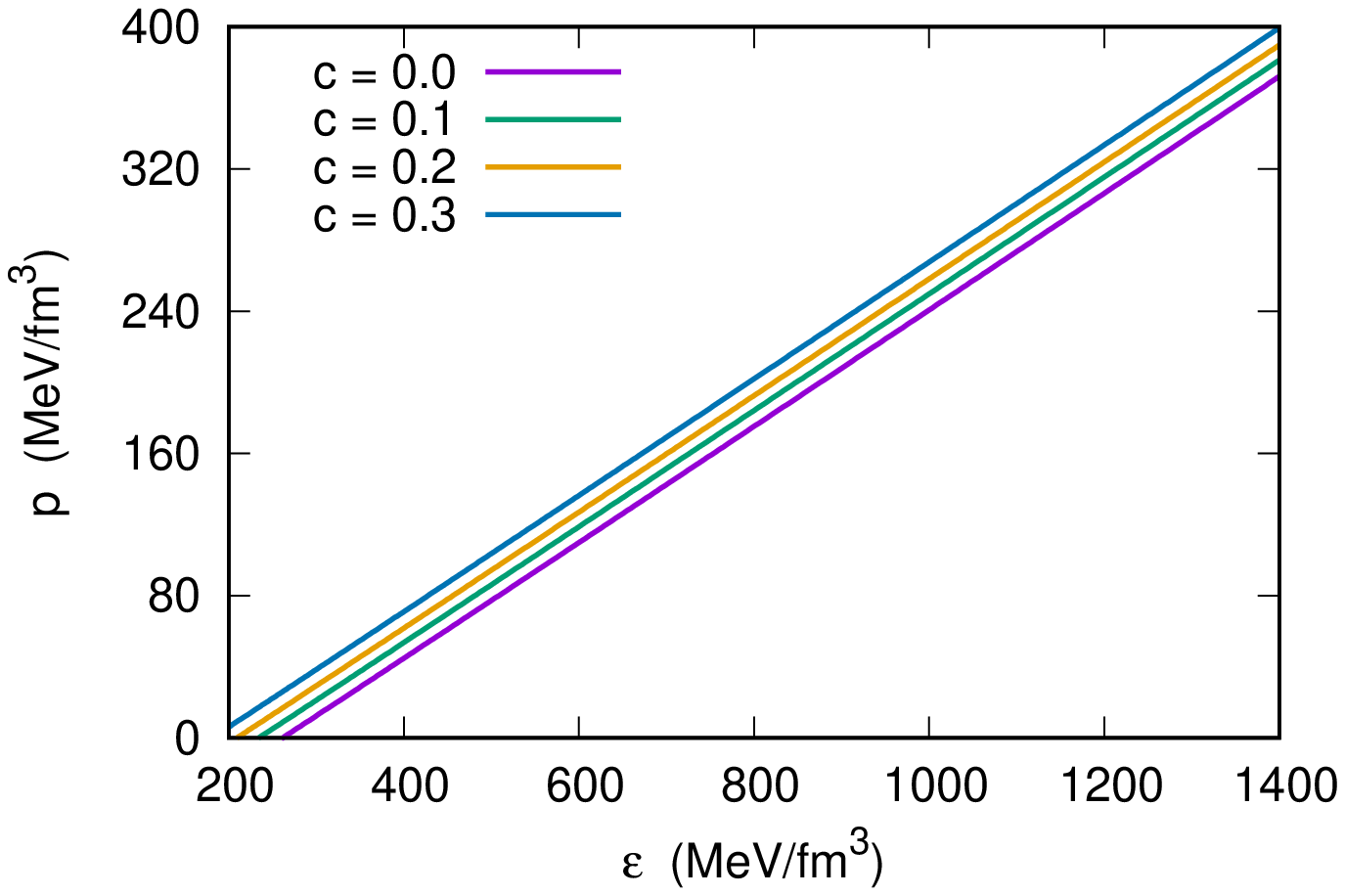} &
\includegraphics[width=5.6cm,height=7.0cm,angle=270]{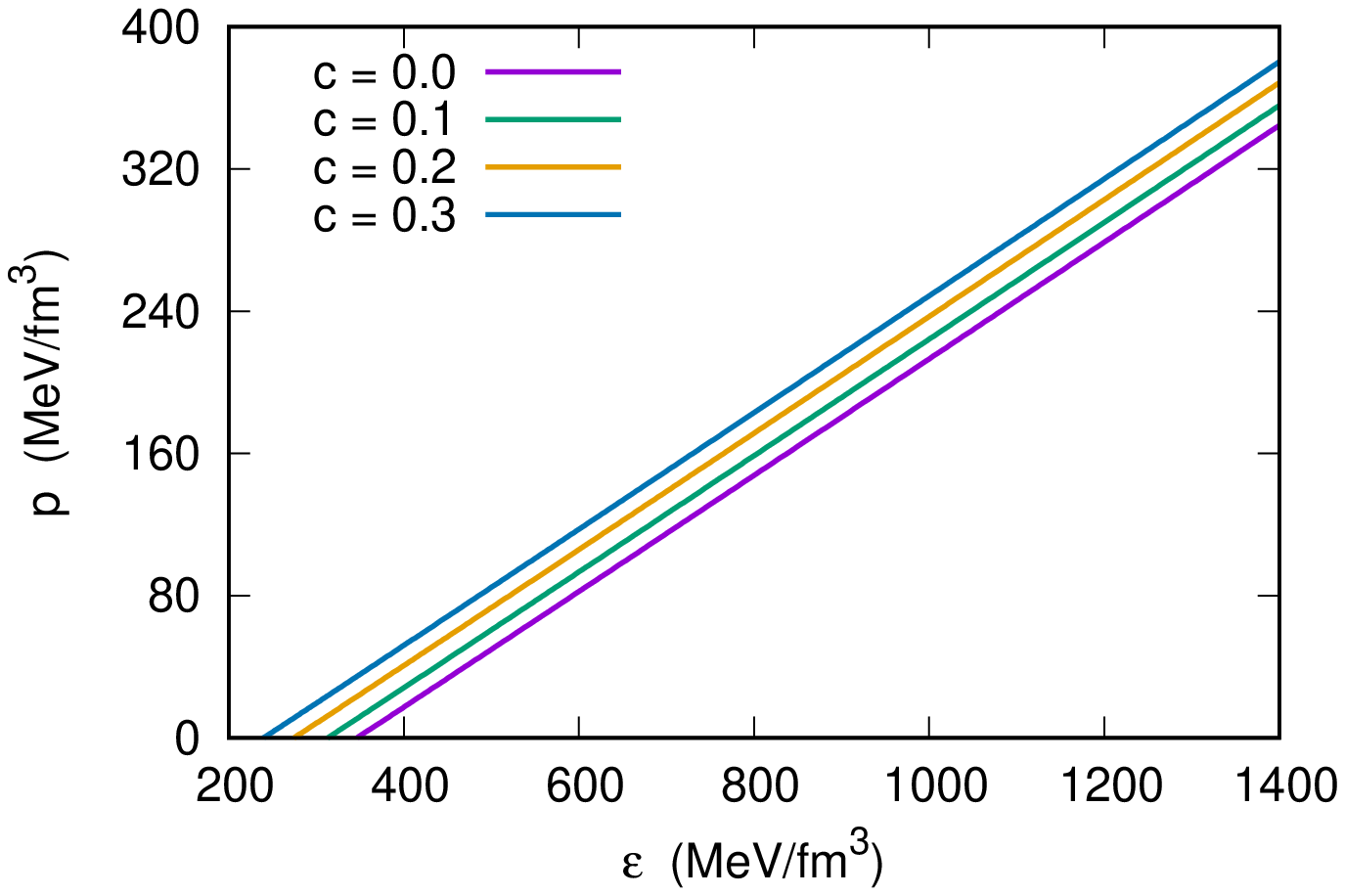} \\
\includegraphics[width=5.6cm,height=7.0cm,angle=270]{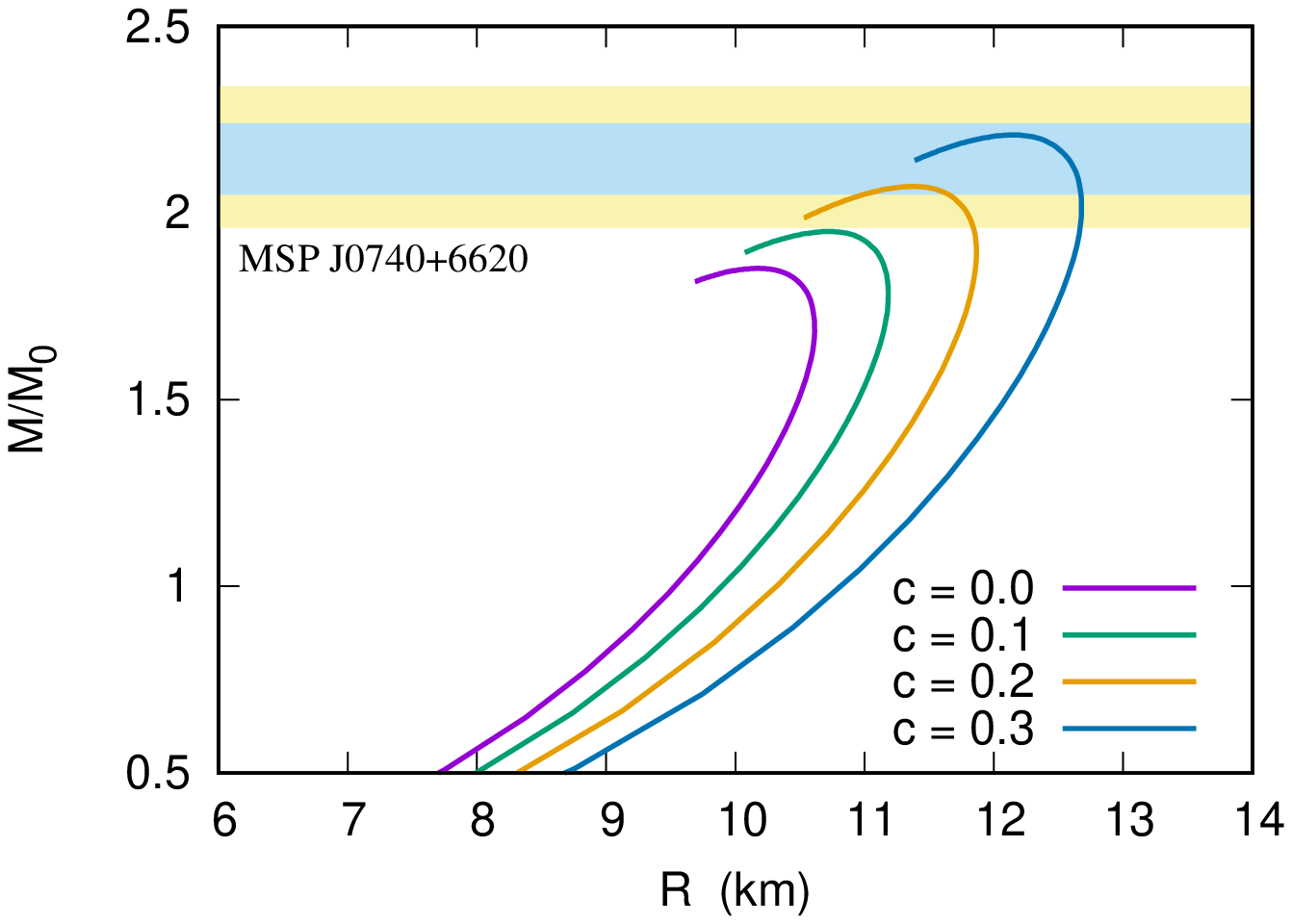} &
\includegraphics[width=5.6cm,height=7.0cm,angle=270]{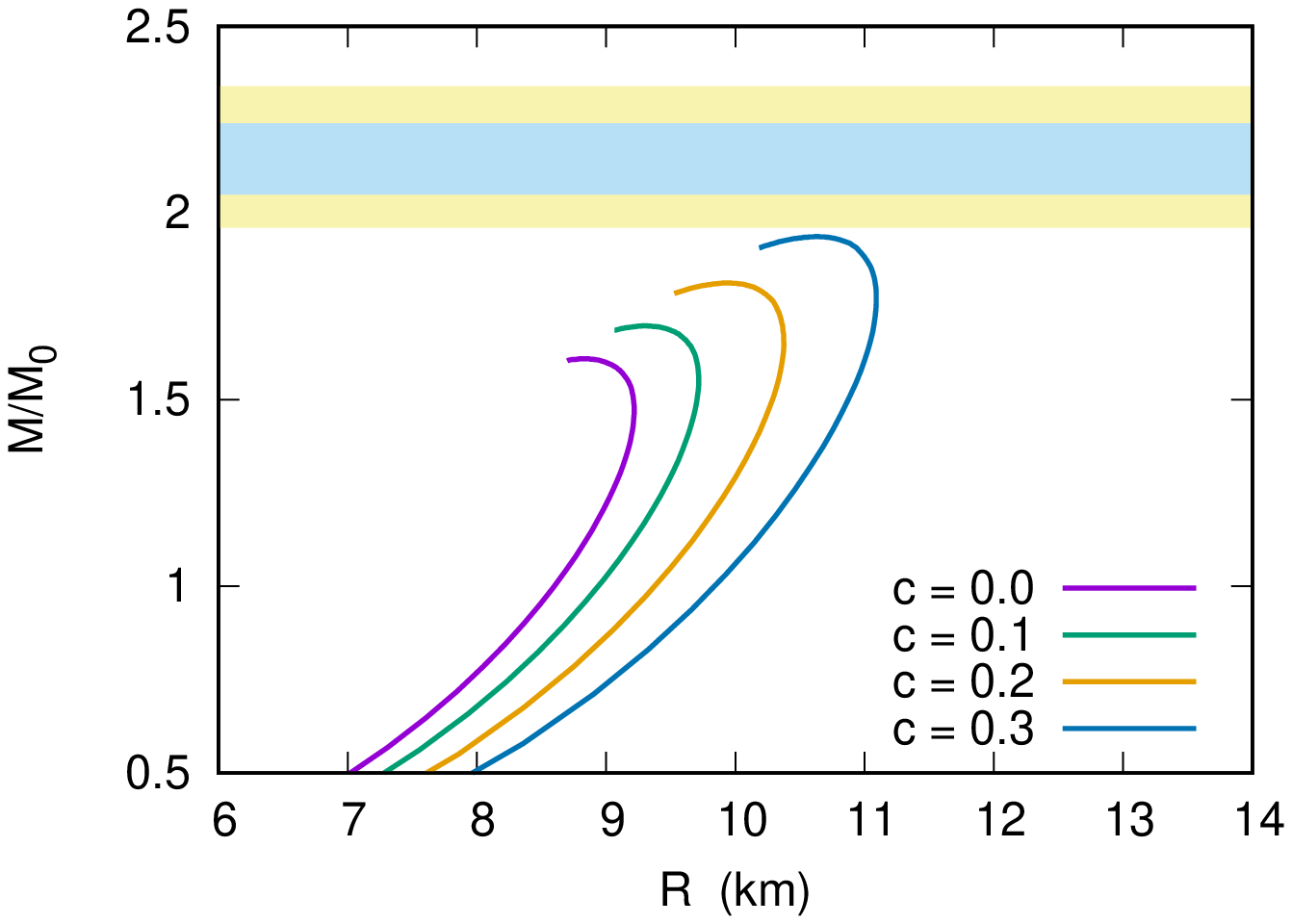} \\
\end{tabular}
\caption{(Color online) EoS (top) and mass-radius relation (bottom) for the minimum (left) and the maximum (right) bag pressure value that produces stable strange stars as a function of the parameter $c$. } \label{F1b}
\end{figure*}
  
  As the $u$ and $d$ quark masses are the same, to avoid saturating the figure, (or display several ones) we plot in Fig.~\ref{F1a} the   strangeness fraction instead of the individual particle population, for  the values of $c$ presented in Tab.~\ref{T2}. The strangeness fraction is defined as:
  
  \begin{equation}
 f_s =  \frac{n_s}{\sum n_f} =  \frac{1}{3} \frac{n_s}{n}, \label{R7}
\end{equation}
where $n$ is the total baryon number density,  $n_s$ is the strange quark number density
and the sum runs over the three quark flavors. As we can see, when we increase the value of $c$ we also increase the strangeness fraction at a fixed density. This was expected once this non-ideal bag model modifies the quark number, but not the chemical potential of the particles.

With the stability window we can now study what the maximum mass of a stable
strange  star is. To accomplish this, we need to construct a neutral, beta-stable 
quark matter. We add leptons as free Fermi gas and impose chemical equilibrium:

\begin{equation}
 \mathcal{L}_{lep} = \sum_l \bar{\psi}_l [i\gamma^\mu\partial_\mu -m_l]\psi_l , \label{R6}
 \end{equation}
 where the sum runs over the two lightest leptons ($e$ and $\mu$), and:
 
  \begin{eqnarray}
  \mu_s =\mu_d = \mu_u + \mu_e  \quad \mbox{and} \quad \mu_e = \mu_\mu , \nonumber \\
  n_s + n_\mu = \frac{1}{3}(2n_u -n_d - n_s). \label{R7}
  \end{eqnarray}

We display in Fig.~\ref{F1b} the EoS and the TOV~\cite{TOV} solution for the minimum and maximum allowed bag pressure values that produce stable strange quark matter. As we increase $c$ we are able to use lower values of the bag pressure. Although we use massive quarks in the EoS, the masses are very low. This makes an almost linear EoS, as in the case of Eq.~(\ref{R2}) for all values of $c$. The EoS differ from each
other only by a displacement proportional to the bag pressure value. Also, lower values of the bag produce more massive quark stars, as pointed in ref.~\cite{Witten,MR2}. Using $c = 0.3$ we are able
 to produce a 2.21 $M_\odot$ stable strange star within the non-ideal bag model, while the original MIT bag model produces a maximum mass of only 1.85 $M_\odot$. Increasing the bag pressure value, the maximum strange quark
star mass decreases, and lies outside the constraint imposed by the MSP J070+6620~\cite{Cromartie}, whose mass is  $2.14^{+0.10}_{-0.09}~M_{\odot}$ at 68$\%$ credibility interval (light blue in Fig.~\ref{F1b}) and $2.14^{+0.20}_{-0.18}~M_{\odot}$ at 95$\%$  credibility interval (light yellow in Fig.~\ref{F1b}). Increasing the bag constant to values beyond the stability window produces unstable strange matter that can be still present in the core of massive hybrid stars. A 
curious feature is the fact that while the central density decreases with the increase of $c$, the strangeness fraction $f_s$ remains almost the same.

Nowadays, an important discussion issue is the radii of canonical stars, $M = 1.4M_\odot$.
The main results of our model are shown in Tab.~\ref{T2a}. In the last decade, several studies pointed towards a small radius. Using X-ray telescopes in quiescent neutron stars, a radius between 10 and 13 km was obtained, as shown in ref.~\cite{Lattimer2013,Catu}. But, here we
pay special attention to a recent study based on multi-messenger observations of the binary neutron-star merger GW170817. The authors~\cite{Capano2020} conclude that the canonical star radius cannot exceed 11.9 km. This result together with the existence of MSP J070+6620 puts strong constraints on the EoS of dense matter. As we can see, for $c =0.2$ and $c = 0.3$ we fulfill both constraints.

However, it must be clear that although the results coming from ref.~\cite{Lattimer2013,Catu,Capano2020} are observational constraints,
they still use some nuclear model to fit the data. Therefore the limit value of 11.9 km must be faced with care.

To summarize this section, we state that we are able to produce massive stable strange stars using the modified MIT bag model. However we have to keep in mind that it is only possible to attain the thermodynamic consistency using a renormalized Fermi-Dirac distribution, which cannot be justified at the level of the simple MIT bag model.

\begin{widetext}
\begin{center}
\begin{table}[ht]
\begin{center}
\begin{tabular}{|c|c|c|c|c|c||c|c|c|c|c|} 
\hline
 ~ c~ &~$M/M_\odot$ - B$_{(Min)}$ ~& R  (km) & $\epsilon_c$  (MeV/fm$^3$)  & $f_s$ & R$_{1.4}$  &~ $M/M_\odot$ - B$_{(Max)}$ & R  (km)&  $\epsilon_c$  (MeV/fm$^3$)  & $f_s$ & R$_{1.4}$  \\
\hline
 0.0    & 1.85 & 10.17& 1286  & 0.322 & 10.38 & 1.61 & 8.93 & 1726 & 0.323 & 9.20     \\
 0.1    & 1.95  & 10.72 & 1158  & 0.322  & 10.81 & 1.70 & 9.31 & 1540  & 0.323 & 9.64   \\
 0.2    & 2.07 & 11.37 & 1038  & 0.321 &11.30 & 1.81 & 9.94 & 1359 & 0.323 & 10.17   \\
 0.3    & 2.21 & 12.18 & 888  & 0.321 & 11.89 & 1.94 & 10.61 & 1205 & 0.322 & 10.74   \\
 \hline
 \end{tabular} 
\caption{Quark stars main properties for different values of $c$. R$_{1.4}$ is given in km.} 
\label{T2a}
\end{center}
\end{table}
\end{center}
\end{widetext}

\section{Vector MIT bag model}

One way to introduce an interaction among the quarks in the MIT model
is by coupling the quarks to a field. We next use a vector field that
produces a repulsion between the quarks.
The inclusion of vector channels in the MIT bag model is not new, 
as can be seen in ref.~\cite{Ro1,Ro2,Ro3,Klan1,Klan2,weiwei}. Following ref.~\cite{Ro2,Ro3},
we introduce the Lagrangian that becomes:

\begin{equation}
\mathcal{L} = \sum_{u,d,s}\{ \bar{\psi}_q  [ \gamma^{\mu} (i\partial_\mu - g_{qqV}V_\mu) - m_q ]\psi_q - B \}\Theta(\bar{\psi_q}\psi_q), \label{v1}
\end{equation} 
where the quark interaction is mediated by the vector channel $V_\mu$
analogous to the $\omega$ meson in QHD ~\cite{Serot}. Indeed, in this
work we consider that the vector channel is the $\omega$ meson itself.

Unfortunately, in the previous papers, the authors missed the mass
term of the vector channel. As we expose below, in mean field approximation,
the vector channel becomes zero if the mass is zero. Therefore, we
introduce the mass term of the $V_{\mu}~(\omega_\mu)$ field as:

\begin{equation}
\mathcal{L}_V = \frac{1}{2}m_V^2V_\mu V^\mu. \label{v2}
\end{equation} 

Now, assuming mean field approximation (MFA)
($V^\mu\rightarrow\langle V \rangle\rightarrow\delta_{0,\mu}V^0$),
we obtain the eigenvalue for the energy of the quarks and the equation of motion for the $V$ field respectively:

\begin{eqnarray}
E_q =  \mu = \sqrt{m_q^2 + k^2} + g_{qqV}V^0, \nonumber \\
m_V^2V_0 =  \sum_{u,d,s}g_{qqV}\langle\bar{\psi_q}\gamma^0\psi_q\rangle , \label{v3}
\end{eqnarray}
where the term $\langle\bar{\psi_q}\gamma^0\psi_q\rangle$ can be
recognized as the number density $n_q$  for each quark $q$.
It is clear from the expression above that if we do not take into account the mass of vector channel, the vector
field itself needs to be zero. The energy density for the quarks is then:

\begin{equation}
\epsilon_q = \frac{N_c}{\pi^2}\int_0^{k_f} E_q k^2 dk. \label{v4}
\end{equation}

 We next need to compute the influence of the massive $\omega$
 particle  on the EoS.  In MFA, we have: $\epsilon = - \langle \mathcal{L} \rangle$. So, the total energy density reads:

\begin{equation}
\epsilon =  \sum \epsilon_q + B - \frac{1}{2}m_V^2V_0^2 , \label{v5}
\end{equation}
the last term of Eq.~(\ref{v5}) being absent in ref.~\cite{Ro1,Ro2,Ro3}. Moreover, the bag vacuum value is not independent of the vector field $V_0$,
which ultimately depends on the strength of the coupling constant. The pressure is obtained via the thermodynamic relation,
$p =  n\mu - \epsilon$, to guarantee thermodynamic consistency
given in Eq.~(\ref{e5}).

To construct a new stability window, we have to fix the coupling
constant $g_{qqV}$, as well the mass of the vector field $m_V$. We consider that the vector channel is the physical $\omega$ meson, 
as in QHD models~\cite{Serot}.  In relation to the coupling constant $g_{qqV}$,
 we have two concerns. The first one refers to its absolute
 value. There are very few studies trying to constrain its value, and
 the uncertainty is yet very high~\cite{MB,kita}.
Most of the models just consider it as a free parameter~\cite{Debora2014,Hana,Shao}.   
Our second concern is about the relative strength of the $g_{qqV}$ constant for different quarks. In the
literature, the $g_{qqV}$'s are universal, assuming the same value for all quark flavors.~\cite{Ro1,Ro2,Ro3,MB,kita,Debora2014,Hana,Shao,Sugano,Lopes2020,lopes2021broken}.
Here we follow a new path. Instead of a universal coupling, we use
symmetry group to fix the relative quark-vector field interaction.  We obtain the relation: 

\begin{equation}
g_{ssV} = \frac{2}{5}g_{uuV} = \frac{2}{5}g_{ddV} .
\end{equation}

All the calculations are detailed in the appendix. The use of symmetry
group to fix coupling constants is very common when we are 
dealing with baryons~\cite{Lopes2020,lopes2021broken,Lopes2014,Dover84,Swart63,Pais,Weiss1,Weiss2,Lopes2015}, but, as far as we known, it is an original
approach in the quark sector. 

In the following, we redefine $(g_{uuV}/m_V)^2 = G_V$ and define $X_V$
as the ratio between $g_{ssV}$ to $g_{uuV}$:

\begin{equation}
 X_V =  \frac{g_{ssV}}{g_{uuV}}. \label{v6}
\end{equation}

\begin{figure}[ht] 
\begin{centering}
 \includegraphics[angle=270,
width=0.4\textwidth]{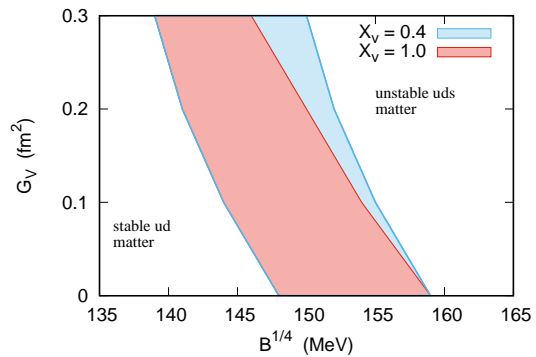}
\caption{(Color online) Stability window obtained with the MIT bag model with
vector interaction and different values of $X_V$.}\label{F2}
\end{centering}
\end{figure}

\begin{table}[h]
\begin{center}
\begin{tabular}{|c|c|c|c|}
\hline
 ~ $G_V~(fm^2)$  ~& $X_V$  &~Min. $B^{1/4}$~&~Max. $B^{1/4}$~ \\
\hline
 0.0 & - &      148 MeV & 159 MeV     \\
 0.1 & 1.0 &    144 MeV & 154 MeV     \\
  0.1 & 0.4 &   144 MeV & 155 MeV     \\
 0.2  & 1.0 &   141 MeV & 150 MeV     \\
  0.2 & 0.4 &   141 MeV & 152 MeV     \\
 0.3  & 1.0 &   139 MeV & 146 MeV     \\
  0.3   & 0.4 & 139 MeV & 150 MeV     \\

 \hline
 \end{tabular} 
\caption{Stability windows obtained with the vector MIT bag model.} 
\label{T3}
\end{center}
\end{table}

\begin{widetext}
\begin{center}
\begin{table}[ht]
\begin{center}
\begin{tabular}{|c|c|c|c|c|c|c||c|c|c|c|c|}
\hline
 ~ $G_V$ $(fm^2)$~ & $X_V$ &~$M/M_\odot$ - B$_{(Min)}$ ~& R  (km) & $\epsilon_c$  (MeV/fm$^3$)  & $f_s$  & R$_{1.4}$ &~ $M/M_\odot$ - B$_{(Max)}$ & R  (km)&  $\epsilon_c$  (MeV/fm$^3$)  & $f_s$   & R$_{1.4}$ \\
\hline
 0.0    & - & 1.85 & 10.17& 1286  & 0.322 & 10.38 & 1.61 & 8.93 & 1726 & 0.323  & 9.20   \\
 \hline
 0.1    & 1.0 &2.16  & 11.29 & 1051  & 0.320 &11.11 & 1.92 & 9.98 & 1331  & 0.321 & 10.08    \\
 0.2    & 1.0 &2.43 & 12.24 & 921  & 0.318 & 11.66 & 2.19 & 11.00 & 1112 & 0.320 & 10.79    \\
 0.3    &  1.0 &2.61 & 12.97 & 795  & 0.317 & 12.08  & 2.40 & 11.85 & 978 & 0.319 & 11.34   \\
 \hline
  0.1    & 0.4 &2.09 & 11.11& 1071  & 0.354 & 11.02 & 1.82 & 9.63 & 1450 & 0.379   & 9.85  \\
  0.2    & 0.4 &2.28 & 11.83 & 962  & 0.379 & 11.50 & 2.01 & 10.36 & 1248 & 0.388 & 10.38     \\
  0.3    &  0.4 &2.41 & 12.33& 893  & 0.402 & 11.81 &  2.14 & 10.86 & 1154 & 0.413  & 10.72   \\
 \hline
 \end{tabular} 
\caption{Quark star main properties for different values of $G_V$ and $X_V$.} 
\label{T3a}
\end{center}
\end{table}
\end{center}
\end{widetext}

We construct a stability window for $G_V$ varying from 0 to 0.3 fm$^2$
for $X_V = 1$, which is usually found in the literature, and for $X_V
= 0.4$, which is the value predicted by symmetry group. The results are
presented in Fig.~\ref{F2}, as well as in Tab.~\ref{T3}.

As we increase the value of $G_V$, the stability window is displaced
to lower values. For $X_V=1.0$, the stronger the vector field, the narrower the stability window. For $X_V=0.4$, even though the window is also displaced to lower values, its interval remains the same, i.e., $11$~MeV, regardless of the value of $G_V$.
As $X_V = 1.0$ produces a stronger repulsion between the quarks, 
it yields a narrower stability window. As expected, the minimum value
of the stability window is independent of $X_V$, and the differences increase as we increase $G_V$.

We  next study stable strange stars within the vector MIT bag model.  As we are interested in massive strange stars, we only display in the figures the results for the minimum allowed bag vacuum value. Nevertheless, the star masses for the maximum values of the bag are presented in the text as well as in Tab.~\ref{T3a}.

\begin{figure*}[ht]
\begin{tabular}{cc}
\includegraphics[width=5.6cm,height=7.0cm,angle=270]{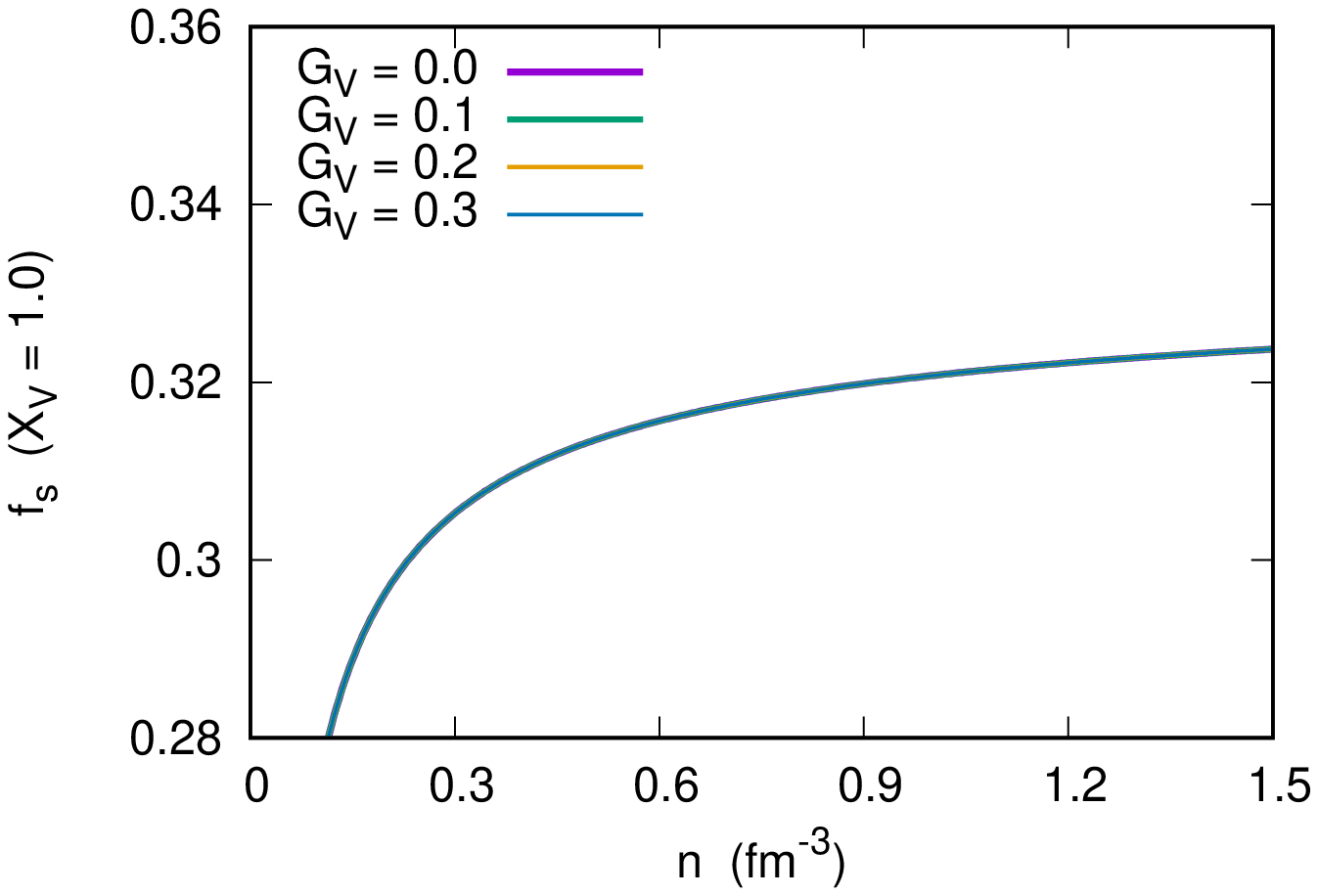} &
\includegraphics[width=5.6cm,height=7.0cm,angle=270]{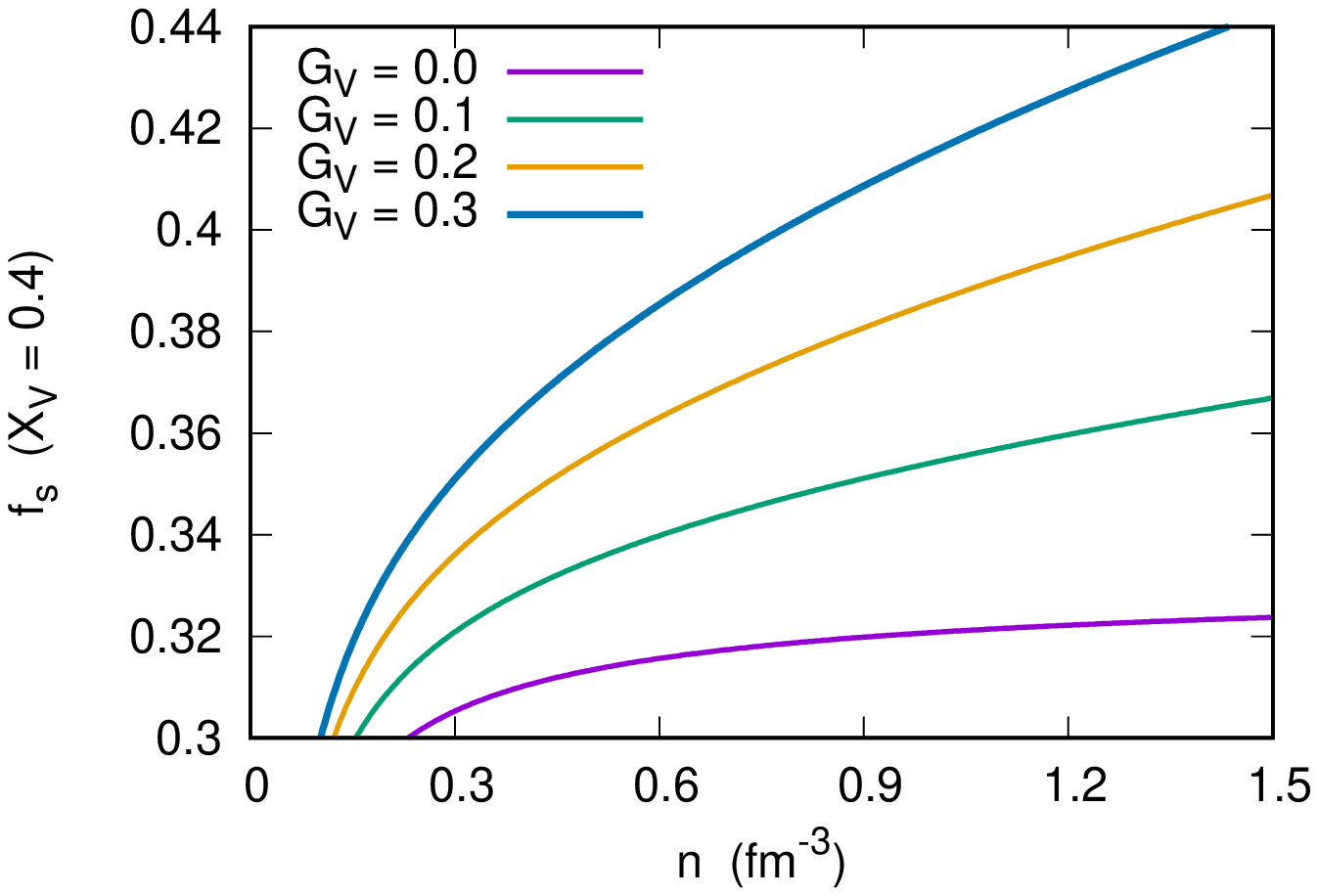} \\
\end{tabular}
\caption{(Color online) Strangeness fraction ($f_s$) as function of $G_V$ for $X_V = 1.0$ (left) and $X_V = 0.4$ (right). } \label{F2b}
\end{figure*}

\begin{figure*}[ht]
\begin{tabular}{cc}
\includegraphics[width=5.6cm,height=7.0cm,angle=270]{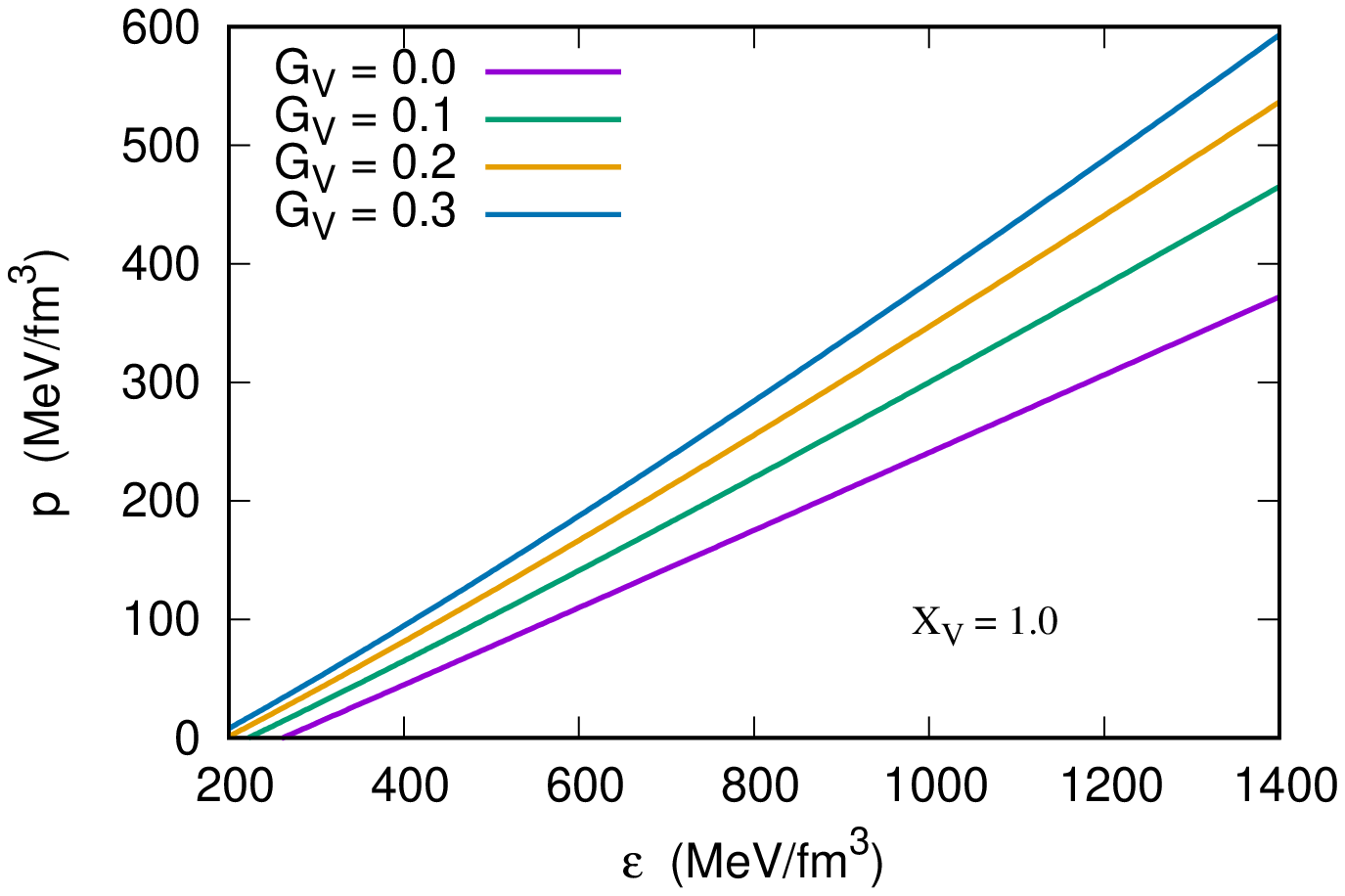} &
\includegraphics[width=5.6cm,height=7.0cm,angle=270]{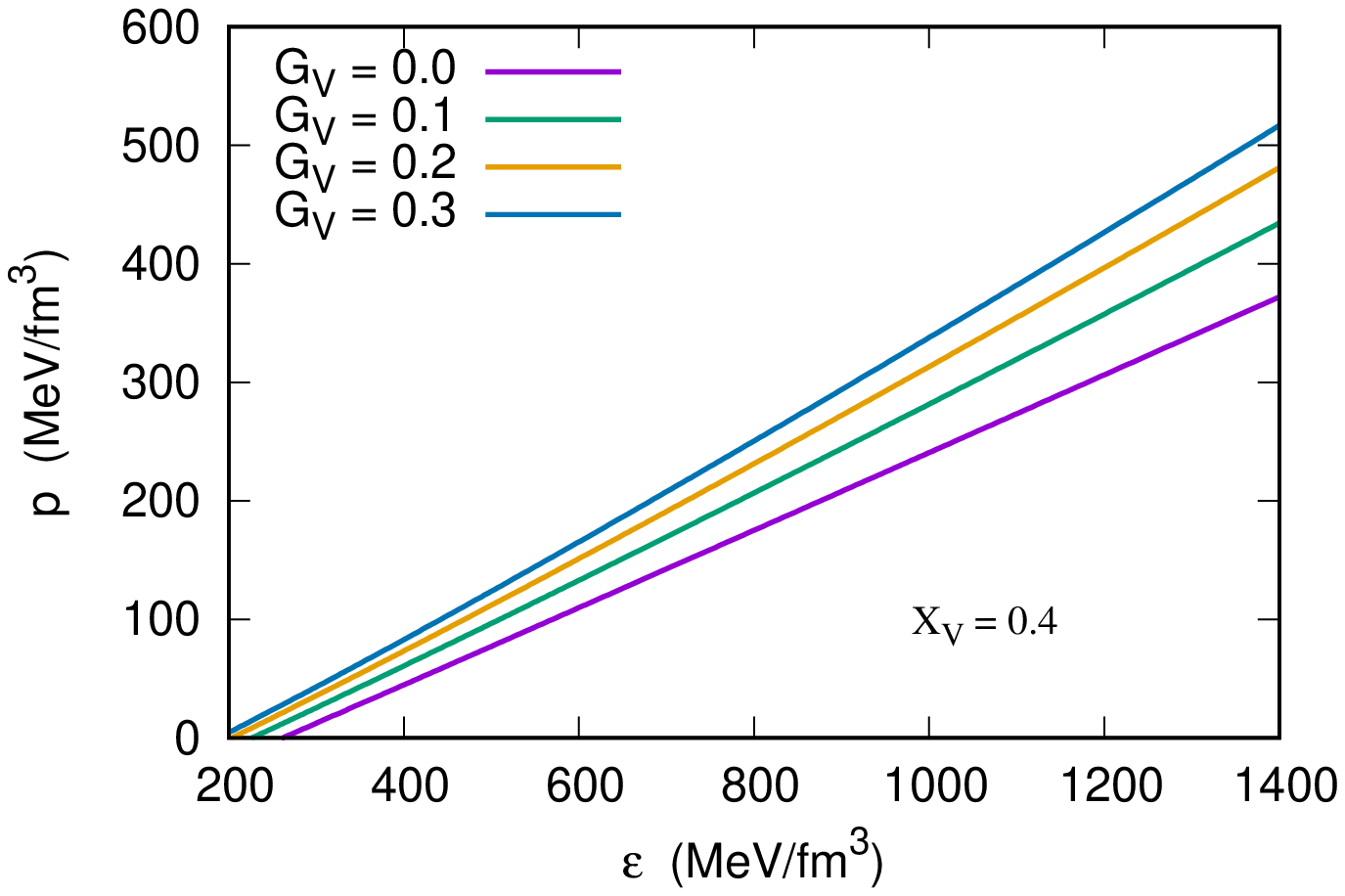} \\
\includegraphics[width=5.6cm,height=7.0cm,angle=270]{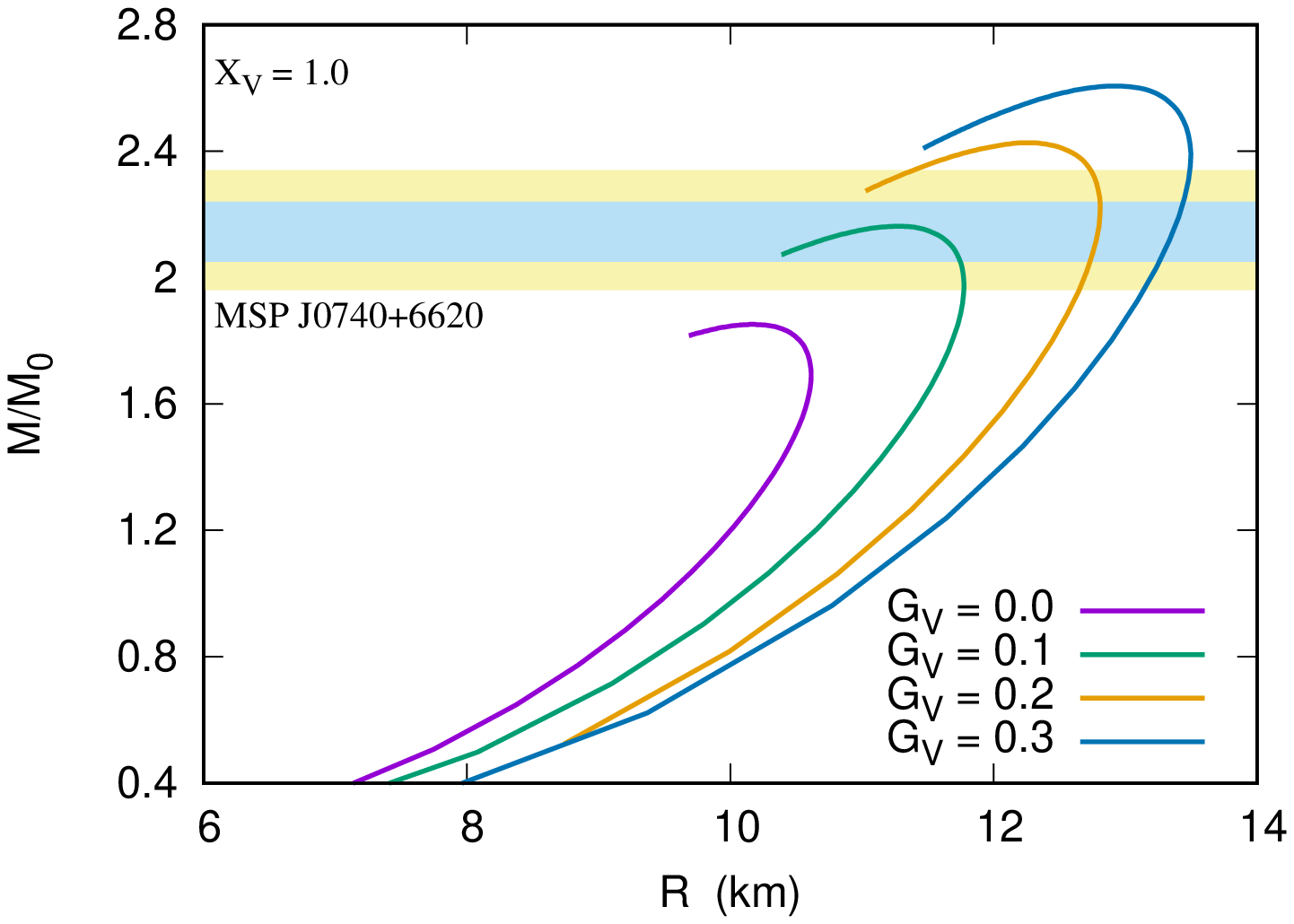} &
\includegraphics[width=5.6cm,height=7.0cm,angle=270]{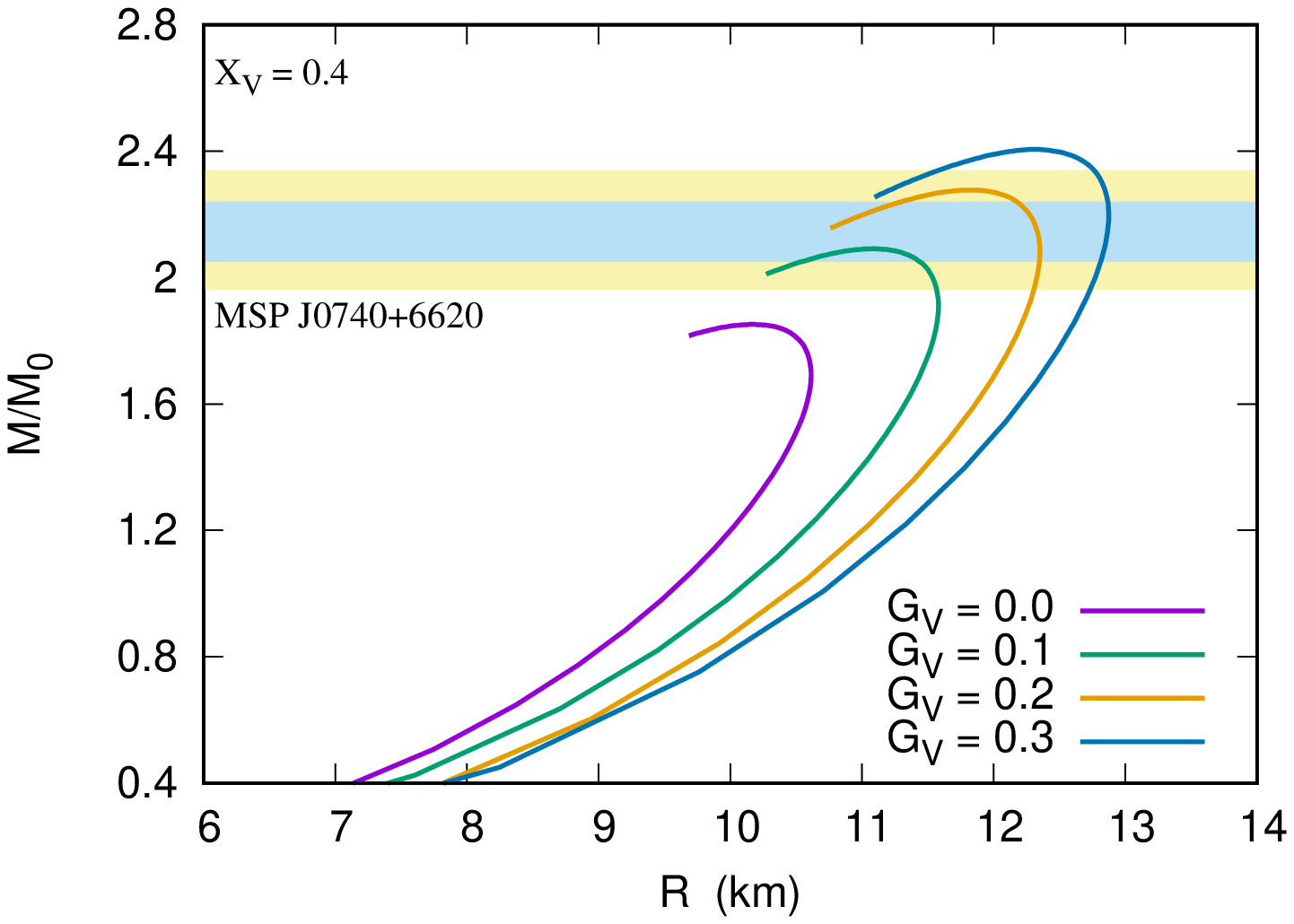} \\
\end{tabular}
\caption{(Color online) EoS (top) and mass-radius relation (bottom) for $X_V = 1.0$ (left) and the $X_V = 0.4$ (right) with the minimum bag pressure value that produces stable strange star as a function of the parameter $G_V$. } \label{F2c}
\end{figure*}

We plot in Fig.~\ref{F2b} the strangeness fraction for different values of $G_V$ and the two different values of $X_V$ previously justified: a universal coupling ratio $X_V = 1.0$ and $X_V = 0.4$. The strangeness fraction is independent of the bag vacuum value. As expected, for the universal coupling, the strangeness fraction is independent of the strength of the coupling of the quarks with the vector field. This can be easily seen from Eq.~(\ref{v3}). As all $g_{qqV}$ have the same values, the chemical potential of the quarks are shifted by the same amount. This behaviour is  similar to the Nambu Jona-Lasinio model for quarks, where the vector field does not affect the particle population either~\cite{Debora2014,Lopes2020, lopes2021broken}.  However, if the couplings are not the same, the $s$ quark, which has the lower value of the coupling constant, has also the lower shift in the chemical potential. As we increase the $G_V$, the difference in the shift becomes larger. 
This causes, for instance, the strangeness fraction at $n = 1.0$ fm$^{-3}$ to grow from 0.32 for $G_V = 0.0$ to 0.415 for $G_V = 0.3$~fm$^2$.

In Fig.~\ref{F2c} we display the EoS and the macroscopic properties of stable strange stars for the minimum allowed value of the bag in the stability window. In this case, we have a {\it positive feedback} between the vector channel and the stability window. As we increase the parameter $G_V$ we stiffen the EoS. Nevertheless, this also reduces the minimum value of the bag pressure in the stability window, which in turn, also stiffens the EoS. The higher the $G_V$ value, the stiffer the EoS. Additionally, the higher the $G_V$, the lower is the minimum allowed value of the bag pressure. The lower the value of the bag pressure, the stiffer the EoS again~\cite{Witten}.
As result of this combined effect we are able to produce very massive stable quark stars, fully compatible with the MSP J0740+6610~\cite{Cromartie}.  Our maximum mass can reach 2.61$M_\odot$ for $X_V =  1.0$. Also, assuming a universal coupling, $X_V = 1.0$, the strangeness fraction at the center of the quark stars decreases as we increase $G_V$ but increases if we assume $X_V = 0.4$. 

Unlike the minimum bag pressure value, the maximum allowed bag pressure value does depend on $X_V$ as can be seen in Fig.~\ref{F2}. As $X_V = 0.4$ produces smaller repulsion, the {\it positive feedback} plays its role again. A small repulsion produces a softer EoS, which produces a higher value of the maximum 
allowed bag pressure value, which in turn also softens the EoS. As a consequence, the maximum quark star mass for $G_V = 0.3$   fm$^2$ can vary from  2.14 $M_\odot$ to 2.61 $M_\odot$, a difference of 0.47 $M_\odot$, i.e. 22$\%$. We emphasize  that our radii are in agreement with ref.~\cite{Capano2020}, except when we assume $B^{1/4}$ = 139 MeV, $X_V = 1.0$, and $G_V$ = 0.3 fm$^2$. This indicates that the bag pressure value is too low, or/and $X_V$ = 0.4 is a better approach to dense quark matter. Anyway, it is clear from Tab.~\ref{T3a} that, within the vector MIT bag model, we are able to produce stable strange stars with masses above 2.4$M_\odot$ that fulfill all astrophysical constrains. 

We can also compare our results with different models. For instance, in ref.~\cite{Klan1}, the authors used a chiral bag NJL-based model with vector interaction, and found that it is not possible to build stable strange matter, even after introducing a deconfinement bag constant. Nevertheless, non-stable strange matter can be important to investigate the possibility of hybrid stars. This is beyond the scope of this work but the reader can seek more information in refs.~\cite{Klan1,Klan2,weiwei}.

In this section we have shown that we are able to obtain very massive quark stars,  all models are thermodynamically consistent and no modifications on the Fermi-Dirac distribution are necessary. Another point worth mentioning: in our work the $G_V$ values are around ten times smaller than the one used in ref.~\cite{Ro2,Ro3} where $G_V = 2.2$ fm$^2$. This is probably because the authors did not include the mass term of the vector field. Only for a  comparison, if we remove the mass term, Eq.~(\ref{v2}), from our final Lagrangian, the maximum mass drops from 2.61 $M_\odot$ to only 2.08 $M_\odot$, a reduction of 0.53 $M_\odot$!

\section{Self-interacting vector field}

The vector channel in mean field approximation takes into account only
the valence quarks. This scenario is called "no sea approximation", once the
Dirac sea of quarks is completely ignored~\cite{Serot97}. As the vector field
is borrowed directly from quantum hadrodynamics (QHD), the vector MIT bag model also becomes renormalizable~\cite{Serot}.
However instead of transforming the mean field approximation (MFA) into a more complex relativistic Hartree or Hartree-Fock approximation, we can take the Dirac sea into account throughout modifications on
the effective Lagrangian as made in ref.~\cite{Serot97}. 
Here, we introduce a quartic contribution for the  vector field: $(V_{\mu}V^\mu)^2$ as a correction
for the EoS at high density which will mimic the Dirac sea contribution.

When we add the vector channel in the MIT bag model Lagrangian, it creates a
repulsion term in the quark-quark interaction and, as result, the pressure (as well as the chemical potential and the energy density) increases. The stiffening of the EoS  grows linear with the density. The quartic vector field makes the EoS more malleable. The introduction of self-interacting fields is not new in the relativistic models. 
 Boguta and Bodmer~\cite{Boguta} introduced self-interaction in the scalar sector to correct the compressibility of the symmetric nuclear matter. The same Bodmer also introduced quartic interaction in the vector  sector~\cite{Bodmer} and others~\cite{Toki1,Toki2} used quartic terms in order to correct the behaviour of nuclear  matter at densities above 2$n_0$. Now we introduce a self-interacting vector field in the vector MIT bag model as:

\begin{equation}
U(V^{\mu}) =   b_4 \frac{(g^2V_\mu V^\mu)^2}{4} \label{nl2} ,
\end{equation}
 where $g = g_{uuV}$ for short, and $b_4$ is a dimensionless parameter~\cite{Serot97}. The self-interactions of the vector field allow us to construct either a softer or a stiffer EoS when compared
with the linear case. It also plays a crucial role in the relation between pressure and chemical potential -$p(\mu)$.
This relation is important in hadron-quark phase transitions~\cite{MR1,MR2,Ro1,Lopes2020}. 
As in the hadronic case~\cite{Toki1}, we do not expect any significant modification in the EoS
for densities below 2$n_0$.
Using a mean field approximation and solving the Euler-Lagrange equations
of motion we obtain the following eigenvalues for the quarks and $V$ field respectively:  

\begin{eqnarray}
E_q =  \mu = \sqrt{m_q^2 + k^2} + g_{qqV}V_0, \nonumber \\
gV_0  + \bigg ( \frac{g}{m_v} \bigg)^2 \bigg (  b_4 (gV_0)^3 \bigg )= \nonumber 
 \bigg (\frac{g}{m_v} \bigg ) \sum_{u,d,s} \bigg (\frac{g_{qqV}}{m_v} \bigg )n_q . \nonumber \\ \label{nl3}
\end{eqnarray}

Again, this result is very similar to the Boguta and Bodmer terms in the scalar and vector fields in hadronic models~\cite{Boguta,Bodmer}. In order to produce only small deviations
we impose that $|b_4| \leq 1$. Due to  the quartic nature of the non-linear term, the higher the density, the higher the
deviation from the linear one. Also, as can be seen from Eq.~(\ref{nl2}), the quartic term has also a quartic dependence on the $g$ constant. So, the higher the value of $G_V$, the higher the influence of the quartic term. The advantage of using non-linear terms is that we can modify the EoS at high density while keeping the stability window unaffected. 

We are now in the position to construct an EoS in MFA, considering the Fermi-Dirac distribution of the quarks and the Hamiltonian of the
vector field and the bag pressure value, $\mathcal{H} = -\langle \mathcal{L} \rangle$. We obtain:

\begin{equation}
\epsilon_q = \frac{N_c}{\pi^2}\int_0^{k_f} E_q k^2 dk. \label{nl4}
\end{equation}

\begin{equation}
\epsilon =  \sum \epsilon_q + B - \frac{1}{2}m_V^2V_0^2  - U(V_0). \label{nl5}
\end{equation}

The pressure is obtained via the thermodynamic relation,
$p =  n\mu - \epsilon$, to guarantee thermodynamic consistency.

\begin{figure*}[ht]
\begin{tabular}{cc}
\includegraphics[width=7.6cm,height=5.8cm,angle=0]{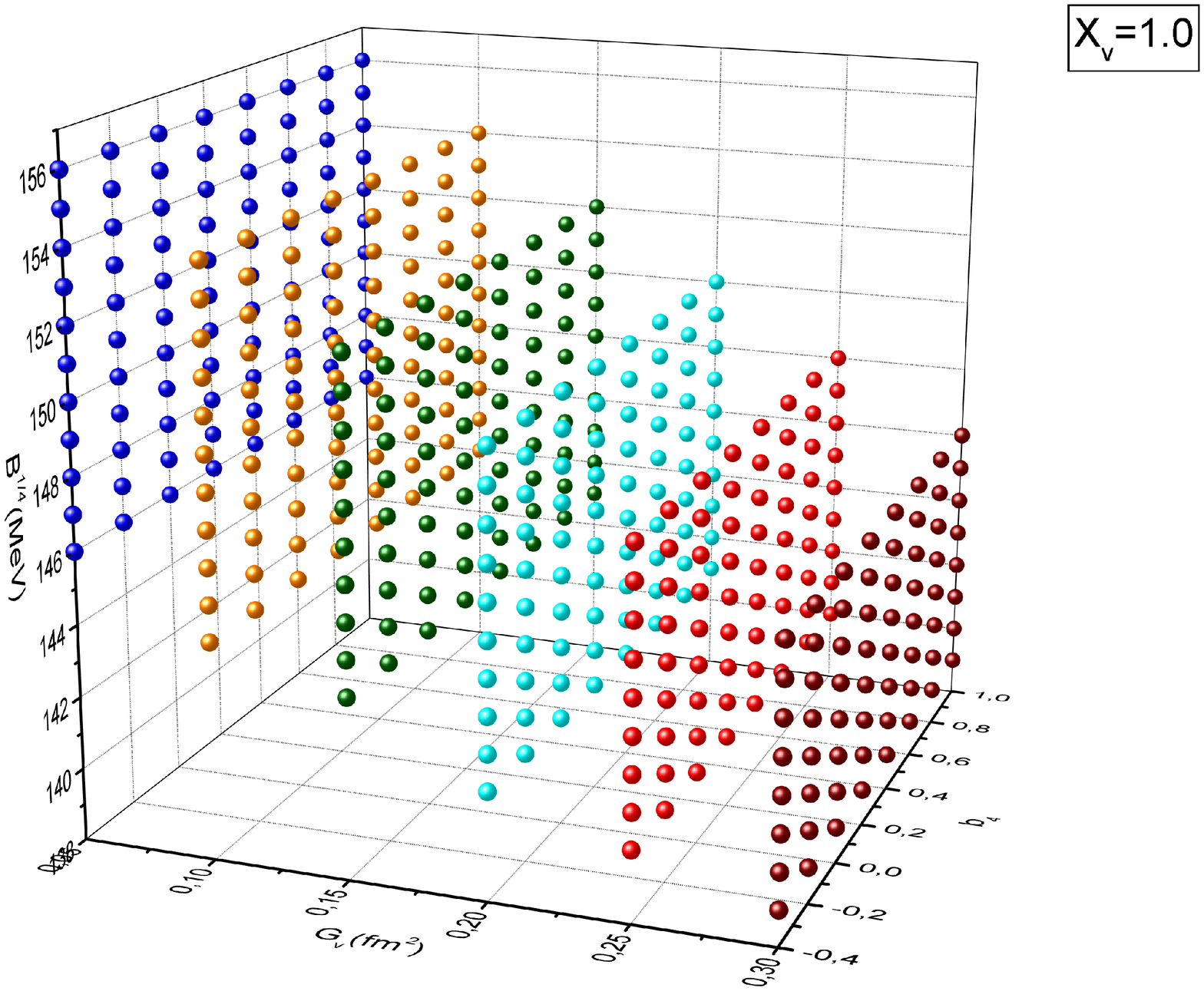} &
\includegraphics[width=7.6cm,height=5.8cm,angle=0]{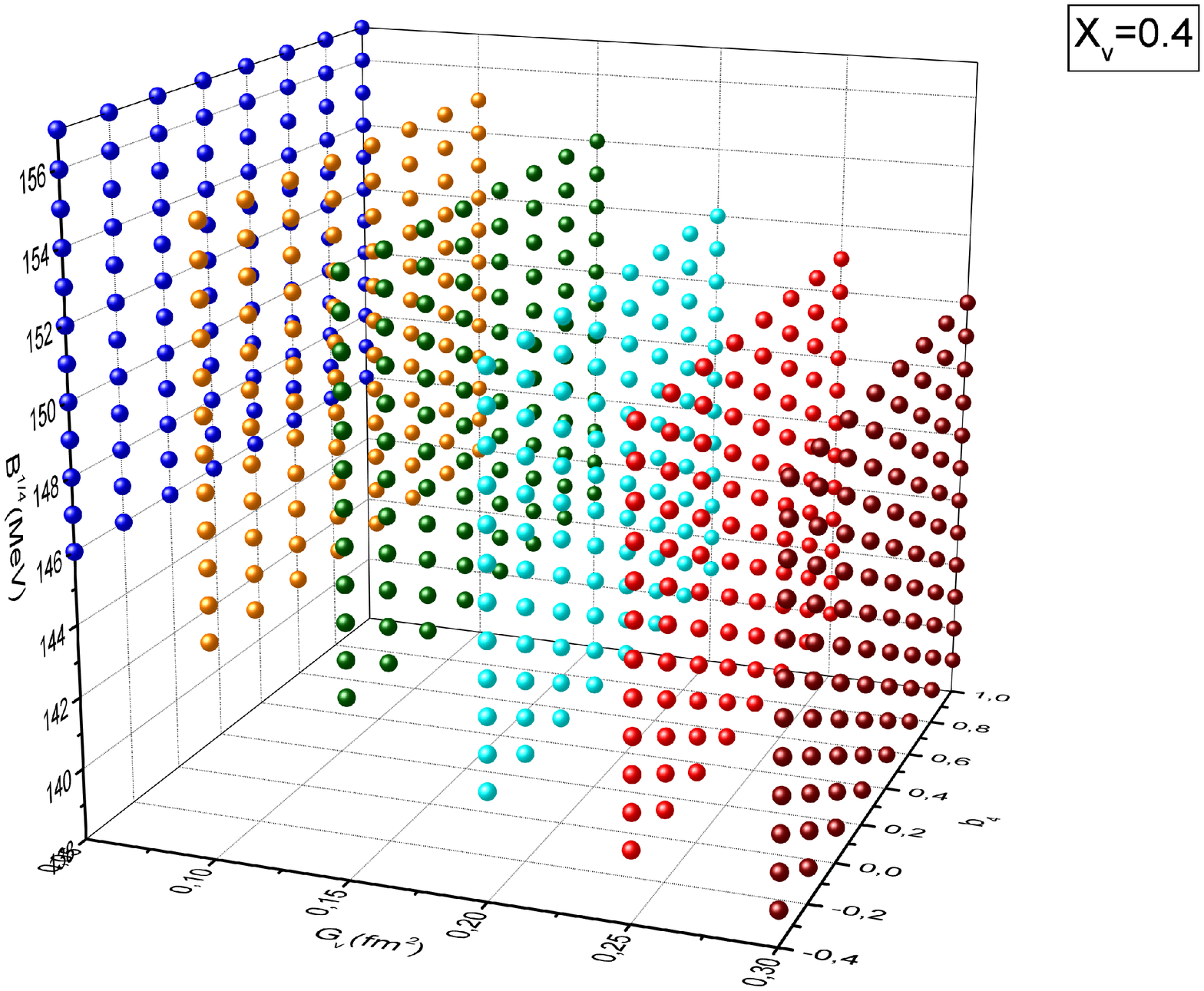} \\
\end{tabular}
\caption{(Color online) 3-D stability wall  for $X_V = 1.0$ (left) and $X_V = 0.4$ (right). Each color wall represent a value of $G_V$. As expected the results are independent of $b_4$. } \label{F3a}
\end{figure*}

In Fig.~\ref{F3a} we display a 3-D stability window. For each value of $G_V$ we vary the 
dimensionless parameter $b_4$ from -0.4 to 1.0.
As pointed out, the advantage of the self-interaction is to 
preserve the original stability window. This can be seen in this 3-D graphic. Each color wall is a stability window for a fixed $G_V$. The results are the same as the ones shown in Tab.~\ref{T3}, being
independent of $b_4$.

\begin{figure}[ht] 
\begin{centering}
 \includegraphics[angle=270,
width=0.4\textwidth]{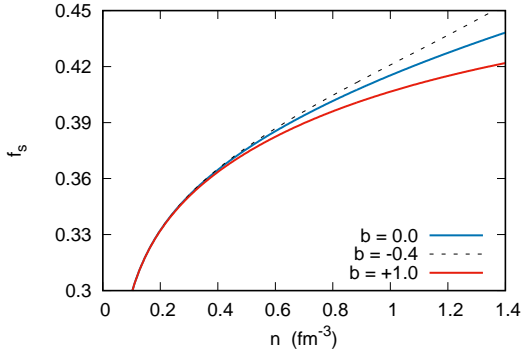}
\caption{(Color online) Strangeness fraction obtained with $G_V = 0.3$ fm$^2$ and $X_V = 0.4$ for different choices of $b$.}\label{F3b}
\end{centering}
\end{figure}

As can be seen from Eq.~(\ref{nl2}), the influence of the self-interaction does not only dependent on 
$b_4$ but also on $g$, as presented in ref.~\cite{Glen} for the scalar meson.
Hence, the higher the value 
of $G_V$, the stronger the influence of the quartic term.  So we display here only the results for $G_V$ = 0.3 fm$^2$. As for lower values, the influence is significantly  lower. We can also increase the strength of $G_V$  or $b_4$, but they would not modify the qualitative aspects and therefore, are beyond the scope of this work.  We start by  studying the influence of the non-linear term in the strangeness fraction $f_s$. If $b_4$ is negative, then the vector field $V^0$ increases with density when compared with the pure linear case.
On the other hand, if $b_4$ is positive, then $V^0$ is lower when compared with the linear case. This effect is also
reflected in the strangeness fraction if $X_V~\neq~1.0$. As already pointed out in the last section, if $X_V = 1.0$ then the strangeness fractions is independent of the vector field. Due to these considerations, here we only discuss the strangeness fraction for $G_V = 0.3$ fm$^2$ and $X_V = 0.4$, as predicted by symmetry group. The results are displayed in Fig.~\ref{F3b}.

As can be seen, the strangeness fraction is the same up to densities about 0.4 fm$^{-3}$. From this point on, the results begin to differ from the linear case. For negative $b_4$ (-0.4), we have 
an increase of the vector field, which makes the differences of the coupling constant more evident,
increasing the fraction of $s$ quarks. On the other hand, a positive $b_4$ (1.0) reduces the vector field as well as the strangeness fraction when compared with the linear case. For instance, fixing $n = 1.0$   fm$^{-3}$ we have $f_s$ equal to 0.415, 0.421, and 0.406 for $b_4$ equal to 0.0, -0.4 and +1.0 respectively. 

\begin{figure*}[ht]
\begin{tabular}{cc}
\includegraphics[width=5.6cm,height=7.0cm,angle=270]{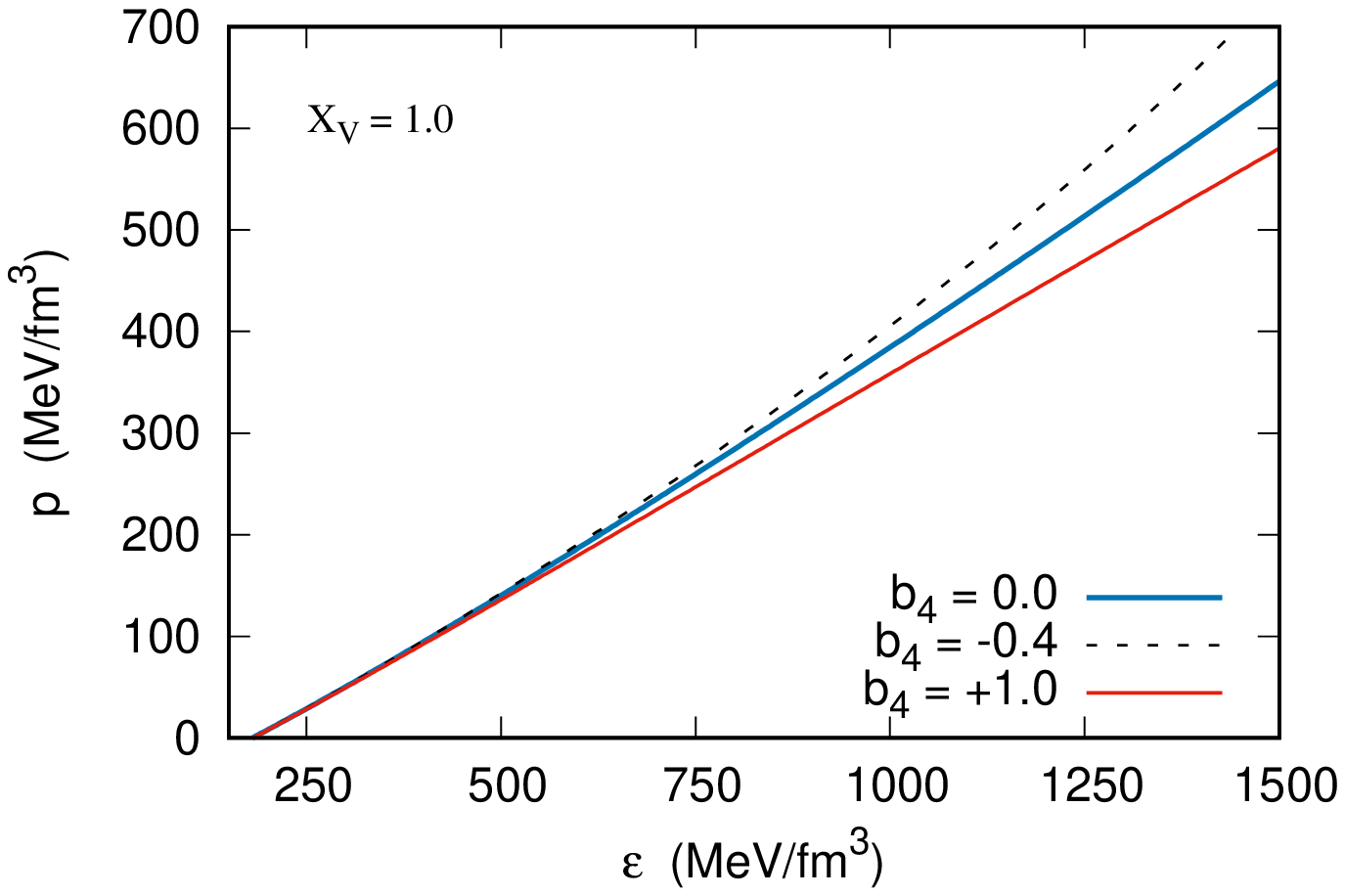} &
\includegraphics[width=5.6cm,height=7.0cm,angle=270]{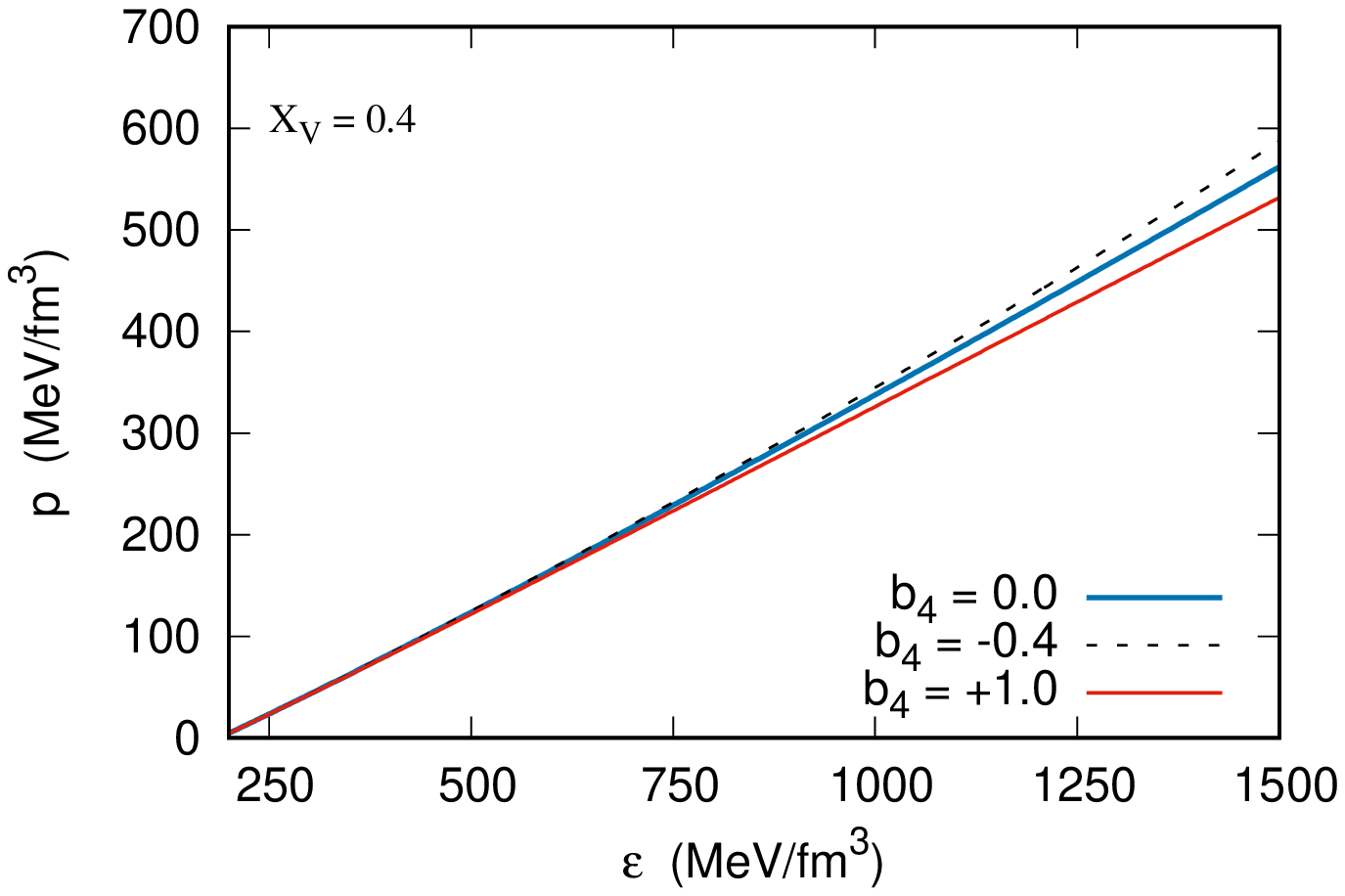} \\
\includegraphics[width=5.6cm,height=7.0cm,angle=270]{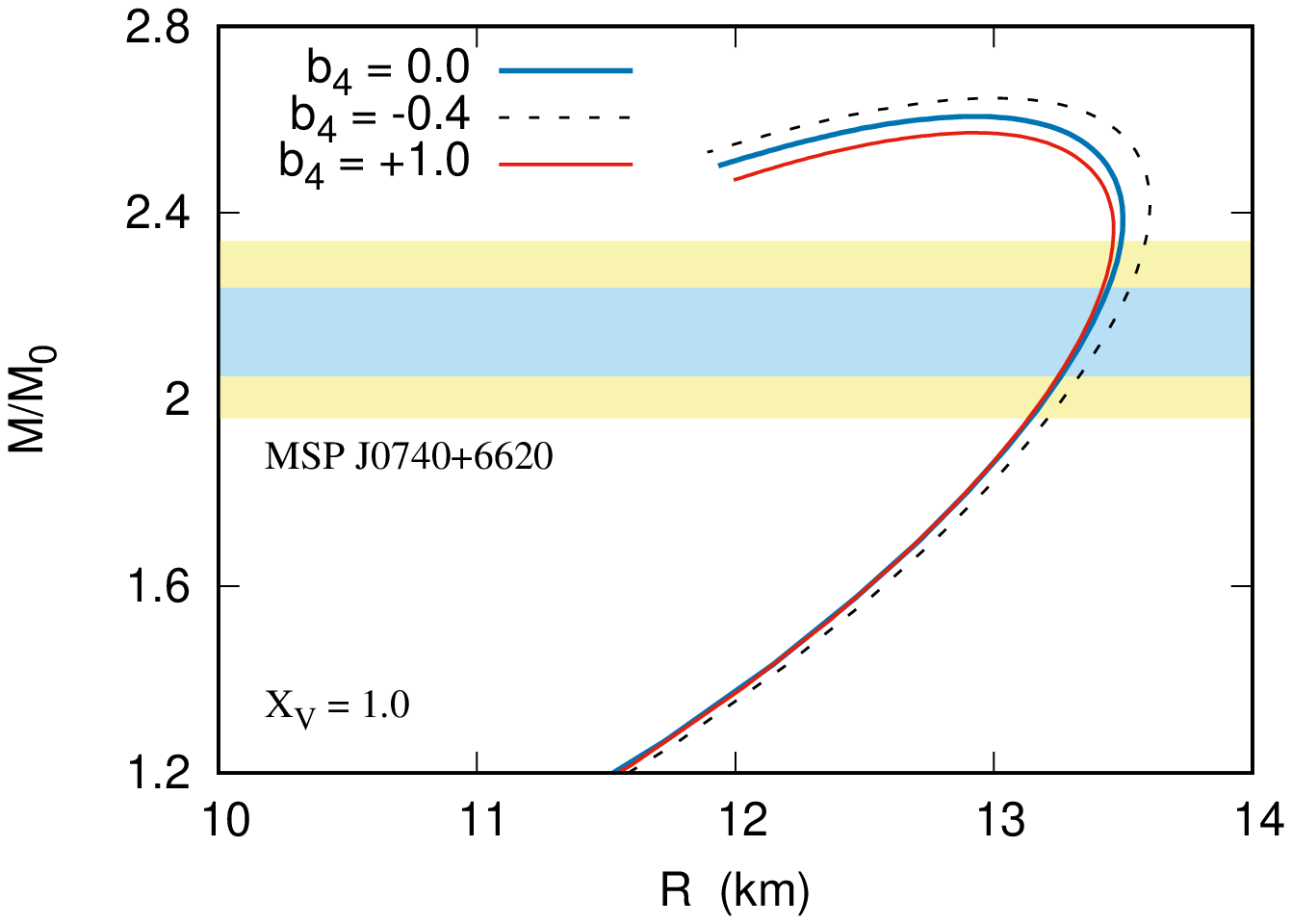} &
\includegraphics[width=5.6cm,height=7.0cm,angle=270]{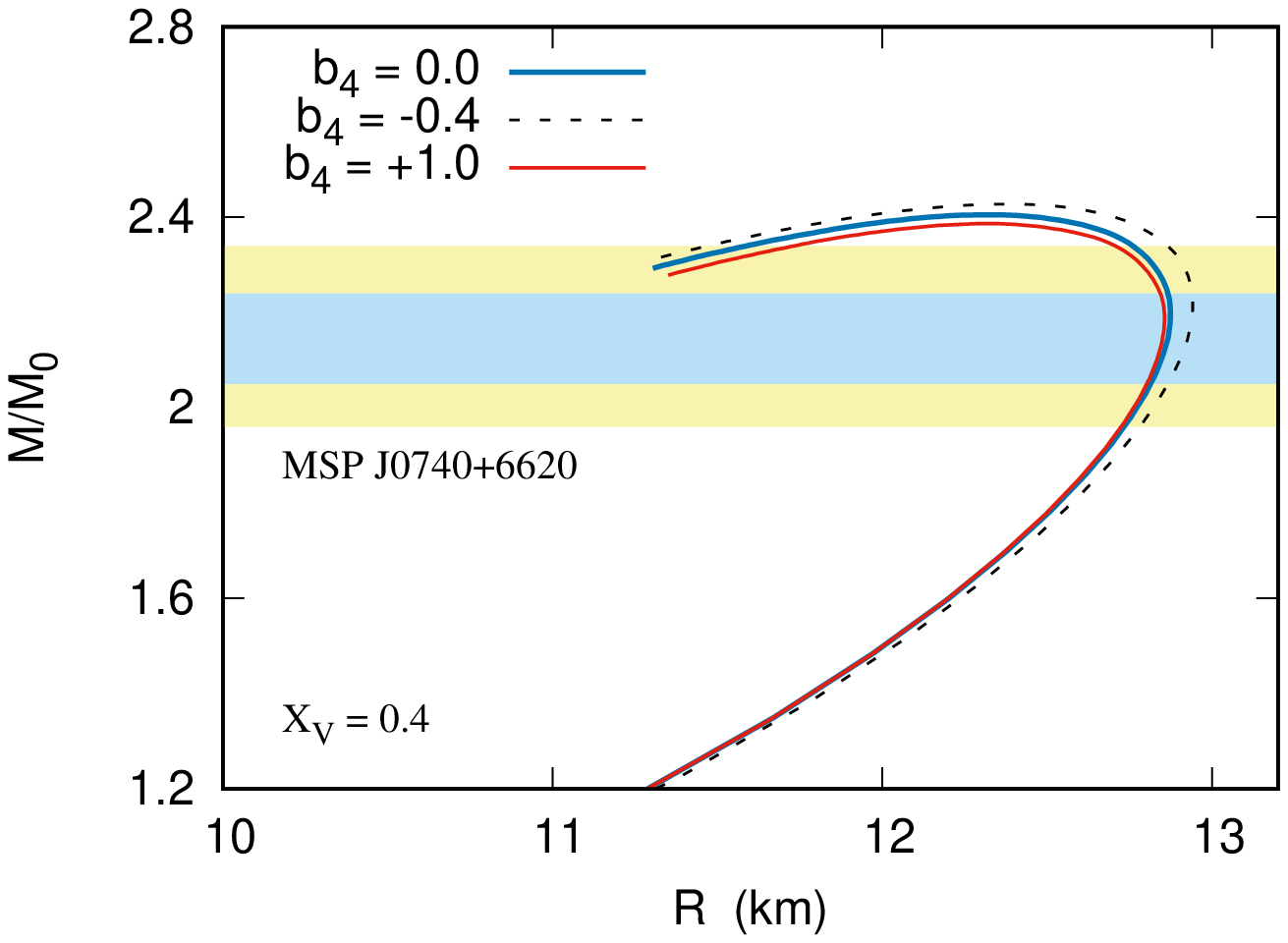} \\
\end{tabular}
\caption{(Color online) EoS (top) and mass-radius relation (bottom) for $X_V = 1.0$ (left) and the $X_V = 0.4$ (right) with the minimum bag pressure value that produces stable strange stars as a function of the parameter $b_4$ and fixed $G_V = 0.3$ fm$^2$. } \label{F3c}
\end{figure*}

We now discuss the effect of the self-interacting vector field on the EoS and on macroscopic properties of strange stars. The results for the minimal allowed bag pressure value for  $G_V =0.3$ fm$^2$ are displayed in Fig.~\ref{F3c}. 

As expected, the EoS is almost the same until the energy density reaches values around 400-500 MeV/fm$^3$. Then, as the strangeness fraction behavior, it starts to deviate from the linear case. 
We also see that the influence of the quartic term is higher for $X_V = 1.0$ as it produces a higher $V^0$ value. The softening of the EoS for a positive value of $b_4$ is analogue to the hadronic case, as shown in ref.~\cite{Toki1,Toki2}. The results of the EoS are reflected in the mass-radius relation.
The maximum quark star mass is higher for $b_4$ = -0.4 and lower for $b_4$ = 1.0. Also, from Fig.~\ref{F3c} and Tab.~\ref{T4a}, we can see that the influence of the quartic term is larger for the maximum bag pressure value when compared with its minimum value. The influence is also stronger for $X_V = 1.0$ when compared with $X_V = 0.4$. For instance, the maximum star mass varies by 0.08$M_\odot$ for $X_V = 1.0$ for the minimum bag pressure value and by 0.10$M_\odot$ for the maximum bag pressure value. For $X_V = 0.4$, the mass variation is 0.04$M_\odot$ for the minimum bag pressure value and 0.06$M_\odot$ for its maximum. The variation in the radius are not significant.
Also, for $X_V = 0.4$ the strangeness fraction at the center of the maximum mass star increases if $b_4$ is negative and decreases if $b_4$ is positive.

We finish this section indicating that although for the specific case that  $B^{1/4}$ = 139 MeV, $G_V = 0.3$ fm$^2$ and $X_V = 1.0$ the radii of canonical stars are in disagreement with ref.~\cite{Capano2020}, in all other cases  we are able to produce  stable strange stars in agreement with this study. Indeed, within the self-interacting vector field, even a 2.44$M_\odot$ star is obtained.

\begin{widetext}
\begin{center}
\begin{table}[ht]
\begin{center}
\begin{tabular}{|c|c|c|c|c|c|c||c|c|c|c|c|}
\hline
 ~ $b_4$ ~ & $X_V$ &~$M/M_\odot$ - B$_{(Min)}$ ~& R  (km) & $\epsilon_c$  (MeV/fm$^3$)  & $f_s$ & R$_{1.4}$  &~ $M/M_\odot$ - B$_{(Max)}$ & R  (km)&  $\epsilon_c$  (MeV/fm$^3$)  & $f_s$ & R$_{1.4}$   \\
\hline
-0.4    & 1.0 &2.65  & 13.02 & 797  & 0.317 &12.13 & 2.44 & 11.85 & 986  & 0.319 &  11.36 \\
 0.0    & 1.0 & 2.61 & 12.97& 795  & 0.317 & 12.08 & 2.40 & 11.85 & 978 & 0.319  & 11.34   \\
 +1.0    & 1.0 &2.57 & 12.96 & 791  & 0.317 & 12.08 & 2.34 & 11.73 & 994 & 0.319 & 11.27     \\
 \hline
  -0.4    & 0.4 &2.43 & 12.45 & 903  & 0.407 & 11.84 & 2.16 & 10.88 & 1188 & 0.417 & 10.88    \\
  0.0    & 0.4 &2.41 & 12.33 & 893  & 0.402 & 11.81 & 2.14 & 10.86 & 1154 & 0.413 & 10.86    \\
  +1.0    &  0.4 &2.39 & 12.35& 869  & 0.396 & 11.80 &  2.10 & 10.77 & 1185 & 0.406 & 10.69     \\
 \hline
 \end{tabular} 
\caption{Quark star main properties for different values of $b_4$ and $X_V$.} 
\label{T4a}
\end{center}
\end{table}
\end{center}
\end{widetext}

\section{Tidal deformability}

Before we finish we make a brief discussion about the tidal deformability $(\Lambda)$
and the recent constraints presented in the literature coming from VIGO and LIGO gravitational
wave observatories and the  GW170817 event~\cite{VL}.

If we put an extended body  in an  inhomogeneous external field it will experience different forces throughout its extent. The result is a tidal interaction.
The tidal deformability of a compact object is a single parameter $\lambda$ (sometimes called Love parameter) that quantifies how
easily the object  is deformed when subjected to an external tidal field. A larger tidal deformability indicates that the object is easily deformable. On
the opposite side, a compact object  with a smaller tidal deformability parameter is smaller, more compact, and it is more difficult to deform it.

The tidal deformability is defined as the ratio between the induced quadrupole
$Q_{i,j}$ and the perturbing tidal field $\mathcal{E}_{i,j}$ that causes the perturbation:

\begin{equation}
 \lambda = -\frac{Q_{i,j}}{\mathcal{E}_{i,j}} .
\end{equation}

However, in the literature is more commonly found the dimensionless tidal deformability parameter $(\Lambda)$ defined as:

\begin{equation}
 \Lambda \doteq \frac{\lambda}{M^5} \equiv \frac{2k_2}{3C^5} ,
\end{equation}
where $M$ is the compact object's mass and $C = M/R$ is its compactness.
The parameter $k_2$ is called Love number and is related to the metric perturbation.
A complete discussion about the Love number and its calculation is both,  very extensive and  well documented in the literature. Therefore, it is out of the scope of the present work. We refer the interested reader to see ref~\cite{Katerina,lopes2021broken,Damour,Poisson,Hin} and the references therein.

\begin{figure}[ht] 
\begin{centering}
 \includegraphics[angle=270,
width=0.49\textwidth]{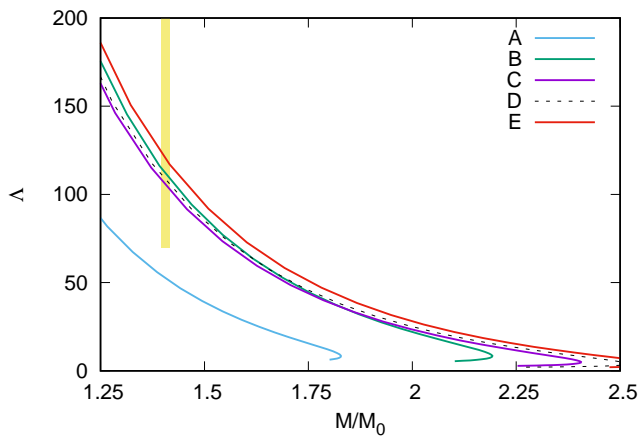}
\caption{(Color online) Dimensionless tidal parameter as a function of the gravitational mass for different modified MIT models. Model A is the original MIT, model B is the non-ideal bag with c = 0.3, model C(D) is the linear vector MIT with $X_V$ = 0.4 (1.0) and model E is the vector MIT with Dirac's sea contribution with $X_V$ = 1.0 and $b_4$ = -0.4. The yellow line is the LIGO/VIRGO constraint $70 < \Lambda_{1.4} < 580$ for the 1.4$M_\odot$ star.}\label{FT}
\end{centering}
\end{figure}

Now we can discuss how the different modifications of the MIT affect its tidal deformability. In all chosen models, we use the minimum bag value to
fulfill Bodmer-Witten conjecture. Also, we use c = 0.3 in the non-ideal bag model,
and $G_V$ = 0.3 fm$^2$ in all vector couplings. The tidal
deformabilities for the canonical star 1.4$M_\odot$ ($\Lambda_{1.4}$) for different models are presented in Tab.~\ref{TF} and the tidal deformabilities for stars up to 2.5$M_\odot$,
as well the VIGO/LIGO constraint ($70 < \Lambda_{1.4} < 580$) provided in by ref.~\cite{VL} are showed in Fig.~\ref{FT}.

\begin{table}[ht]
\begin{center}
\begin{tabular}{|c|c|c|c|c|c|c|}
\hline
 - & A  & B &  C  & D & E\\
\hline
$\Lambda_{1.4}$  &  54  & 113 & 107  & 111 & 122     \\
 
 \hline
 \end{tabular} 
\caption{Dimensionless tidal parameter for the canonical mass ($\Lambda_{1.4}$) for different models discussed in the text. Only the original MIT (model A) does not fulfill the LIGO/VIRGO constraint.} 
\label{TF}
\end{center}
\end{table}

Before we proceed, its good to remark that, although ref.~\cite{VL} uses
data of the LIGO/VIRGO observatories, the authors still use some of nuclear 
model to fix the parameters. Therefore, as in the radius constraint, the tidal deformability constraint also must be faced with care.

We can see that, with the exception of the original MIT bag model, all discussed
models fulfill the LIGO/VIRGO constraint. Indeed, the values for $\Lambda_{1.4}$ are
all lower than 290, the most probable value discussed in ref.~\cite{VL}. This is the opposite to what normally happens with common used models of neutron stars, which in general present a higher
value for the $\Lambda_{1.4}$ as showed in ref.~\cite{lopes2021broken}. Even the model E,
which has a radius beyond 11.9 presents a relatively small tidal deformability.
This reflects the fact that the Love number $k_2$ has a higher dependence on
the microscopic EoS than the macroscopic mass and radius as pointed out in ref.~\cite{Hin}.

\section{Conclusions}

In this work we revisited the MIT bag model, as well as some of its modified versions found in the literature~\cite{MR1,MR2,Ro1,Ro2,Ro3}. We started from the Lagrangian density and obtained the stability window for the original model. Then we revisited the so called non-ideal bag model, obtained the stability windows for some values of the parameter $c$. To preserve the thermodynamic consistency, the Fermi-Dirac statistics needs to be modified. In fact, the only modification required is in the quark number distribution. Once it is modified, the correction in the energy density and pressure comes as consequence. Therefore this model modifies the statistics rather than the MIT bag model itself. 
A stable strange star of mass as high as 2.21$M_\odot$ and a radius of 11.89 km, which agrees with both mass and radius constraints~\cite{Cromartie,Capano2020}, were produced. However, we must remember of using
a renormalized Fermi-Dirac distribution.

We can overcome the thermodynamic issues by introducing a vector field in the Lagrangian density. We first corrected this introduction, as originally proposed in ref.~\cite{Ro1,Ro2,Ro3,Klan1}, by taking into account the mass term of the vector field. With the corrected Lagrangian density we reduced the value of $G_V$ from 2.2 fm$^2$ to values ten times smaller. Moreover, besides the traditional universal coupling for all three quarks, with the help of symmetry group arguments, we calculated a new coupling constant for the $s$-quark mass, which is 40$\%$ of the $u$ and $d$ quark couplings.   We constructed a stability window for this model and we were able to reproduce a star with a mass of 2.61$M_\odot$, but in disagreement with the maximum radius
obtained in ref.~\cite{Capano2020}. Yet, we were able to produce a 2.41$M_\odot$ that fulfils all the
astrophysical constraints while keeping the thermodynamic consistency. 

At the end we proposed a modification on the linear vector field, inspired in the models of QHD~\cite{Boguta,Bodmer,Toki1,Toki2}, by including a quartic term in the vector channel. We emphasize here that this is meant to be only a correction in order to mimic the contribution of the Dirac quark sea, which is absent in MFA. This quartic term allowed us to modify the EoS while keeping the stability window unaltered. We saw that a positive quartic term contribution causes a softening of the EoS at high densities, while a negative one causes its stiffening. Ultimately a stable strange star with mass of 2.44$M_\odot$ that fulfils the above mentioned astrophysical constraints was obtained. 

To finish our analyses, we calculated the tidal deformability of quark stars. We saw that the canonical star (with 1.4$ M_\odot$), if faced as a quark star, presents a very small tidal deformability ($\Lambda_{1.4})$. The constraint obtained from GW170817 is not very restrictive and allows values between 70 and 580.
In all of our models, this tidal deformability is lower than 130, and lies outside the observational constraint for the original MIT model, with the chosen bag parameter.

 In the second part of this study we investigate the influence of the temperature on the vector MIT bag model, and also check if the model is able to reproduce results coming from both LQCD and from the experimental determination of the chemical freeze-out. 

$ $

{\bf Acknowledgments} 

This work is a part of the project INCT-FNA Proc. No. 464898/2014-5. D.P.M. was partially supported by Conselho Nacional de Desenvolvimento Científico e Tecnológico  (CNPq/Brazil) 
under grant 301155.2017-8  and  C.B. acknowledges a doctorate scholarship from Coordenação de Aperfeiçoamento de Pessoal do Ensino Superior (Capes/Brazil).

\appendix

\section{ SU(3) and SU(6) symmetry group}

To fix the quark coupling constant to the vector channel we use the hybrid group SU(6), which 
is invariant under both SU(3) flavor symmetry and SU(2) spin symmetry. 
We start from the Yukawa Lagrangian~\cite{Swart63}:

\begin{equation}
\mathcal{L} =  -g(\bar{\psi_q}\psi_q)M , \label{a1}
\end{equation}
where $\psi_q$ is the quark Dirac field, and $M$ is the field of an arbitrary meson.
This Lagrangian belongs to the irreducible representation IR$\{1\}$, a unitary singlet.
The $u-d-s$ quarks belong to the IR$\{3\}$ = D(1,0), while the anti-quarks belong
to the IR$\{3^{*}\}$ = D(0,1)~\cite{Swart63}. The meson field can
belong either to IR$\{8\}$ or to IR$\{1\}$.
The direct product of $\{3\}~\otimes~\{3^{*}\}$ = $\{8\}~\oplus~\{1\}$~\cite{Lipkin}. So, to
 preserve the IR$\{1\}$ of the Lagrangian, the ($\bar{\psi_q}\psi_q$) needs to belong to the
IR$\{8\}$ if the meson $M$ belongs to IR$\{8\}$, or belong to IR$\{1\}$ if the meson
belongs to IR$\{1\}$. The coupling constant for each quark can be written as~\cite{Lopes2014}:

\begin{equation}
\mathcal{L} = -g_2^8\mathcal{C}(\bar{\psi_q}\psi_q)M , \label{a2}
\end{equation}
for the mesons belonging to IR$\{8\}$. If the mesons belong to IR$\{1\}$  we have:

\begin{equation}
\mathcal{L} = -g_1(\bar{\psi_q}\psi_q)M , \label{a3}
\end{equation}
where the $\mathcal{C}$ is the SU(3) Clebsch-Gordan (CG) coefficient~\cite{Swart63}.
We calculate the $CG$ with the help of the algorithm presented in ref.~\cite{CGA} and the 
tables presented in ref.~\cite{TCG}

As the quarks are in IR$\{3\}$ and the anti-quarks in IR$\{3*\}$; in $\{3\}~\otimes~\{3^{*}\}$ we have only one IR$\{8\}$ as result, in opposition to 
the baryon case, where the baryons and anti-baryons are in
IR$\{8\}$. In $\{8\}~\otimes~\{8\}$ there are two possible $\{8\}$, typically called symmetric and anti-symmetric ones. This also implies that
the $\alpha_v$ defined in ref.~\cite{Swart63} is always equal to 1 in our case.
The coupling constants read:

\begin{eqnarray}
g_{uu\omega_8} = \bigg (\frac{1}{\sqrt{6}} \bigg) \times \bigg ( \frac{1}{\sqrt{8}} \bigg ) =  g_8 \frac{1}{\sqrt{48}} \nonumber , \\
g_{dd\omega_8} =  \bigg (-\frac{1}{\sqrt{6}} \bigg) \times \bigg ( -\frac{1}{\sqrt{8}} \bigg ) =  g_8 \frac{1}{\sqrt{48}} ,  \label{a4} \\
g_{ss\omega_8} =  \bigg (-\frac{2}{\sqrt{6}} \bigg) \times \bigg ( \frac{1}{\sqrt{8}} \bigg ) =  -g_8 \frac{2}{\sqrt{48}} , \nonumber \\
g_{uu\phi_1} = g_{dd\phi_1} = g_{ss\phi_1}  = g_1. \nonumber
\end{eqnarray}

The coupling of the strange quark is twice the value of the non-strange ones, and  has the opposite sign of the $u-u$ and $d-d$ coupling.
This result is exactly the same as the  $\Sigma-\Sigma$ when compared with the N-N one
 (the reader can consult the table in ref.~\cite{TCG}). Nevertheless,
 as happens in the baryon octet case,  this weird value will be washed out when we impose the
mix of the singlet and octet states.

In nature, the observed $\omega$ and $\phi$ meson are not the theoretical $\omega_8$ and $\phi_1$ but a mixture of them~\cite{Lopes2014,Dover84}.
So, the coupling constant of the real vector mesons with the quarks now reads:

\begin{eqnarray}
g_{uu\omega} = g_{dd\omega}  = g_1 \cos\theta +  g_8 \frac{1}{\sqrt{48}} \sin\theta   , \nonumber \\
g_{ss\omega} =   g_1 \cos\theta -  g_8\frac{2}{\sqrt{48}} \sin\theta . \label{a5}
\end{eqnarray}

Now, to eliminate the last of the free parameters, we impose SU(6) symmetry group, which give us an ideal mixing angle,
($\theta$ = 35.264), and $g_8 =  \sqrt{6}g_1$~\cite{Lopes2014,Dover84,Pais}. We finally obtain: 

\begin{equation}
g_{ss\omega} = \frac{2}{5}g_{uu\omega} = \frac{2}{5}g_{dd\omega} . \label{a6}
\end{equation}

\end{document}